\documentclass[preprint]{aastex}
\usepackage[numberedappendix]{emulateapj5}
\usepackage{epsf}

\slugcomment{To appear in The Astrophysical Journal}

\newcommand\kms{\mbox{km s$^{-1}$}}
\newcommand\Hunits{\mbox{km s$^{-1}$ Mpc$^{-1}$}}
\newcommand\kpc{\mbox{kpc}}
\newcommand\hkpc{\mbox{$h^{-1}\,$kpc}}

\newcommand\hMsun{\mbox{$h^{-1}\,M_\odot$}}
\newcommand\avg[1]{\langle{#1}\rangle}

\newcommand\F{{\cal F}}
\newcommand\lens{{\rm lens}}
\newcommand\src{{\rm src}}
\newcommand\obs{{\rm obs}}
\newcommand\fcool{f_{cool}}
\newcommand\fbar{f_{bar}}
\newcommand\fodd{f_{odd}}
\newcommand\CC{{\hat C}}
\newcommand\M{{\cal M}}
\newcommand\xx{{\bf x}}
\newcommand\uu{{\bf u}}

\newcommand\refeq[1]{eq.~(\ref{eq:#1})}
\newcommand\refEq[1]{Eq.~(\ref{eq:#1})}
\newcommand\refeqs[2]{eqs.~(\ref{eq:#1}) and (\ref{eq:#2})}
\newcommand\refEqs[2]{Eqs.~(\ref{eq:#1}) and (\ref{eq:#2})}
\newcommand\reffig[1]{Figure~\ref{fig:#1}}
\newcommand\reffigs[2]{Figures~\ref{fig:#1} and \ref{fig:#2}}

\makeatletter
\newenvironment{figureone}
  {\def\@captype{figure}}
  {}
\newenvironment{tableone}
  {\def\@captype{table}}
  {}
\makeatother

\begin{document}

\title{Cold Dark Matter and Strong Gravitational Lensing: \\
  Concord or Conflict?}
\lefthead{CDM AND STRONG LENSING}
\righthead{KEETON}
\author{Charles R.\ Keeton}
\affil{Steward Observatory, University of Arizona \\
  933 N.\ Cherry Ave., Tucson, AZ 85721}

\begin{abstract}
Using the number and sizes of observed gravitational lenses, I
derive upper limits on the dark matter content of elliptical
galaxies. On average, dark matter can account for no more than
33\% of the total mass within one effective radius ($R_e$) of
elliptical galaxies, or 40\% of the mass within $2 R_e$ (95\%
confidence upper limits). I show that galaxies built from Cold
Dark Matter (CDM) mass distributions are too concentrated to
comfortably satisfy these limits; a high-density ($\Omega_M=1$)
CDM cosmology is ruled out at better than 95\% confidence, while
a low-density, flat cosmology is only marginally consistent with
the lens data. Thus, lensing adds to the evidence from spiral
galaxy dynamics that CDM mass distributions are too concentrated
on kiloparsec scales to agree with real galaxies, and extends
the argument to elliptical galaxies.

Lensing also provides a unique probe of the very inner regions
of galaxies, because images are predicted to form near the
centers of lens galaxies but are not observed. The lack of
central images in deep maps of radio lenses places strong lower
limits on the central densities of galaxies. The central densities
of CDM galaxies are {\it too low\/} on $\sim$10 parsec scales.
Supermassive black holes can help suppress central images, but
they must lie well off the observed black hole--bulge mass
correlation in order to satisfy current limits on central images.
Self-interacting dark matter, or any other modification to
regular cold dark matter, must simultaneously reduce the
densities on kiloparsec scales and increase the densities on
parsec scales in order to satisfy the unique constraints from
lensing.
\end{abstract}

\section{Introduction}

Cuspy mass distributions are a robust prediction of the popular
Cold Dark Matter (CDM) paradigm. Numerical simulations of
collisionless cold dark matter predict density profiles with
$\rho \propto r^{-\alpha}$ and $\alpha \simeq 1.0$--1.5 at small
radii (e.g., Navarro, Frenk \& White 1996, 1997; Moore et al.\
1998, 1999; Jing \& Suto 2000; Klypin et al.\ 2000), and this
prediction does not depend on particular cosmogonies or initial
conditions (Huss, Jain \& Steinmetz 1999a, 1999b) or on the
specific form of the dark matter power spectrum (Eke, Navarro \&
Steinmetz 2000). Adding dissipative baryons makes mass
distributions even more concentrated (e.g., Blumenthal et al.\
1986; Dubinski 1994). It is important to compare the predicted
mass distributions with real galaxies to test the CDM paradigm
as the explanation for the formation and growth of structure in
the universe.

The dynamics of spiral galaxies have provided the most extensive
observational tests of CDM. Salucci (2001; also see Salucci \&
Burkert 2000, and references therein) suggests that normal spiral
galaxies have dark matter halos with large constant-density cores
as opposed to cusps.  Debattista \& Sellwood (1998) and Weiner,
Sellwood \& Williams (2001) argue that fast-rotating bars require
dark matter densities lower than predicted by CDM.  The slowly-rising
rotation curves of dwarf galaxies and low surface brightness galaxies
also seem to imply constant-density cores (e.g., Flores \& Primack
1994; Moore 1994; McGaugh \& de Blok 1998; Blais-Ouellette, Amram
\& Carignan 2001; de Blok et al.\ 2001). However, several authors
have argued that the H{\sc i} rotation curves may have appeared
artificially shallow due to beam smearing, and that better data
are actually consistent with cuspy CDM mass distributions (van den
Bosch et al.\ 2000; van den Bosch \& Swaters 2000; Swaters, Madore
\& Trewhella 2000).  In rebuttal, McGaugh \& de Blok (1998) and
de Blok et al.\ (2001) claim that beam smearing does not affect
their conclusions; and  Moore (2001) argues that only two of the
19 galaxies analyzed by van den Bosch \& Swaters (2000) are
actually consistent with CDM.  To summarize, many argue that CDM
mass distributions are too concentrated to agree with the dynamics
of spiral galaxies, but the conclusions are still subject to some
vigorous debate.

Other tests of CDM mass distributions should not be affected by
beam smearing. Navarro \& Steinmetz (2000) and Eke et al.\ (2000)
consider the global dynamical properties of spiral galaxies in
terms of the Tully--Fisher (TF) relation between luminosity and
circular velocity. They find that CDM can reproduce the slope and
scatter of the TF relation, but has some trouble with the zero
point. In a high-density ($\Omega_M=1$) cosmology, CDM model
galaxies are too concentrated to agree with the TF zero point,
while in a low-density cosmology the models are marginally
consistent with the data. Taking an entirely different approach,
Rix et al.\ (1997) study the line-of-sight velocity profile of
the elliptical galaxy NGC 2434 using detailed dynamical models.
They find that the galaxy is consistent with CDM models, but the 
strength of the conclusion is limited by systematic uncertainties
such as the orbital anisotropy. These tests do not indicate a
fundamental problem with CDM mass distributions, but they do not
strongly favor the CDM models, either.

The dynamical tests are fundamentally limited by the need to
interpret data from luminosity distributions before drawing
conclusions about mass distributions. Given the importance of
the CDM paradigm in modern cosmology, it is desirable to develop
additional tests that are independent of, and hopefully less
ambiguous than, the dynamical tests. One excellent possibility
is gravitational lensing, because it offers a direct probe of
mass distributions. Individual gravitational lenses robustly
determine the masses of individual galaxies, and the statistical
properties of the lens sample constrain properties of the galaxy
population (e.g., Maoz \& Rix 1993; Kochanek 1993, 1995, 1996;
Cohn et al.\ 2001). Individual lenses and lens statistics both
imply that elliptical galaxies (which dominate the mass-selected
sample of lens galaxies; e.g., Kochanek et al.\ 2000) have
approximately isothermal mass distributions out to several
kiloparsecs, in agreement with the evidence from dynamics (e.g.,
Rix et al.\ 1997) and X-ray elliptical galaxies (e.g., Fabbiano
1989).

It is not clear whether the mass distributions implied by lensing
are consistent with the predictions of CDM. Most lensing studies,
even those that consider a wide range of density profiles (e.g.,
Kochanek 1995; Barkana 1998; Chae, Khersonsky \& Turnshek 1998;
Cohn et al.\ 2001; Mu\~noz, Kochanek \& Keeton 2001), consider
only single-component mass models. However, lens galaxies are
likely to have at least two components (a stellar galaxy and a
dark matter halo), which may contribute comparable amounts of
mass (e.g., Rix et al.\ 1997). Both components are necessary to
explain the distribution of lensed image separations and the fact
that galaxies are much better lenses than more massive groups of
galaxies (Keeton 1998; Porciani \& Madau 2000; Kochanek \& White
2001). Only by allowing two components can we use lensing to
directly test whether real galaxies are consistent with CDM mass
distributions.

The goal of this paper is to use two-component star+halo models
for lensing to test the CDM paradigm. The focus is on elliptical
galaxies because they dominate lens statistics. The outline of the
paper is as follows. Section 2 defines the models, and Section 3
reviews the lensing calculations. In Section 4, the observed number
and sizes of lenses are used to evaluate the global properties of
the models. In Section 5, lensing is used to examine the very
inner regions of galaxies. Finally, Section 6 offers a discussion
and conclusions.

\section{Star+Halo Models}

This section defines star+halo models for elliptical galaxies in
the context of the CDM paradigm. Section 2.1 discusses models for
the stellar and dark matter components, and Section 2.2 gives
normalizations for the models. Only spherical models are considered,
because they are sufficient for calculations of the number and
sizes of lenses. Departures from spherical symmetry mainly affect
the relative numbers of 2-image and 4-image lenses (see Keeton,
Kochanek \& Seljak 1997; Rusin \& Tegmark 2001).

\subsection{Model components}

A simple model for the stellar components of elliptical galaxies is
the Hernquist (1990) model, which has a density profile
\begin{equation} \label{eq:hern}
  \rho(r) = {\rho_s \over {(r/r_s)(1+r/r_s)^3} }\ .
\end{equation}
This profile is described by a scale radius $r_s$, but the
projected profiles of elliptical galaxies are usually described by
an effective (or half-mass) radius $R_e$; they are related by $r_s
= 0.551 R_e$. The total mass of a Hernquist model is $M = 2\pi
\rho_s r_s^3$. The projected surface mass density, in units of the
critical density for lensing, is
\begin{equation}
  \kappa(R) = {\Sigma(R) \over \Sigma_{cr}}
  = \kappa_s\,{ (2+x^2) {\cal F}(x) - 3 \over (x^2-1)^2 }\ ,
\end{equation}
where $x = R/r_s$, $\kappa_s = \rho_s\,r_s / \Sigma_{cr}$, and the
function ${\cal F}(x)$ is:
\begin{equation} \label{eq:F}
  {\cal F}(x) = \cases{
    {1 \over \sqrt{x^2-1}}\,\mbox{tan}^{-1} \sqrt{ x^2-1 } & $(x>1)$ \cr
    {1 \over \sqrt{1-x^2}}\,\mbox{tanh}^{-1}\sqrt{ 1-x^2 } & $(x<1)$ \cr
    1                                                      & $(x=1)$ \cr
  }
\end{equation}
The critical density is $\Sigma_{cr} = (c^2 D_s) / (4 \pi G D_l
D_{ls})$ where $D_l$ and $D_s$ are angular diameter distances to
the lens and source, respectively, and $D_{ls}$ is the angular
diameter distance from the lens to the source (e.g., Schneider,
Ehlers \& Falco 1992). The gravitational deflection for a Hernquist
model is (see eq.~\ref{eq:def1} below)
\begin{equation}
  \phi_{R}(R) = 2\,\kappa_s\,r_s\,{ {x [1-{\cal F}(x)]} \over {x^2-1} }\ .
\end{equation}

Navarro, Frenk \& White (1996, 1997, hereafter NFW) have argued
that dark matter halos found in cosmological $N$-body simulation
of collisionless dark matter have a ``universal'' density profile
of the form
\begin{equation} \label{eq:nfw}
  \rho(r) = {\rho_s \over {(r/r_s)(1+r/r_s)^2} }\ ,
\end{equation}
where $r_s$ is a scale radius and $\rho_s$ is a characteristic
density. It is convenient to replace the scale radius with a
``concentration'' parameter $C = r_{200}/r_s$, where $r_{200}$ is
the radius within which the mean density of the halo is 200 times
the critical density of the universe, which is often taken to mark
the boundary of a relaxed halo (e.g., Crone, Evrard \& Richstone
1994; Cole \& Lacey 1996; Navarro et al.\ 1996, 1997). The
characteristic density is then
\begin{equation}
  \rho_s = {200 \over 3}\,\rho_{crit}(z)\,{ C^3 \over
    {\ln(1+C)-C/(1+C)} }\ ,
\end{equation}
where $\rho_{crit}(z)$ is the critical density of the universe at
the redshift of the halo. The lensing properties of an NFW model
are given by Bartelmann (1996).

More recently, Moore et al.\ (1998, 1999; also see Jing \& Suto
2000; Klypin et al.\ 2000) have argued that the central regions of
simulated halos are steeper than the NFW profile. They advocate a
density of the form
\begin{equation} \label{eq:moore}
  \rho(r) = { \rho_s \over {(r/r_s)^{1.5} [1+(r/r_s)^{1.5}]} }\ .
\end{equation}
For Moore halos, I define a concentration parameter $C =
r_{200}/r_{(-2)}$ in terms of the radius $r_{(-2)}$ at which the
logarithmic slope of the density is $-2$. This definition is
equivalent to the definition of the concentration for NFW halos,
and Keeton \& Madau (2001; also see Wyithe, Turner \& Spergel
2000) argue that it is the best generalization of the
concentration. The radius $r_{(-2)}$ is related to the scale
radius $r_s$ in \refeq{moore} by $r_{(-2)} = 0.630 r_s$. The
characteristic density of a Moore halo is
\begin{equation}
  \rho_s = 25\,\rho_{crit}(z)\,{ C^3 \over \ln[1+C^{3/2}/2] }\ .
\end{equation}

\begin{figure*}[t]
\centerline{\epsfxsize=7.0in \epsfbox{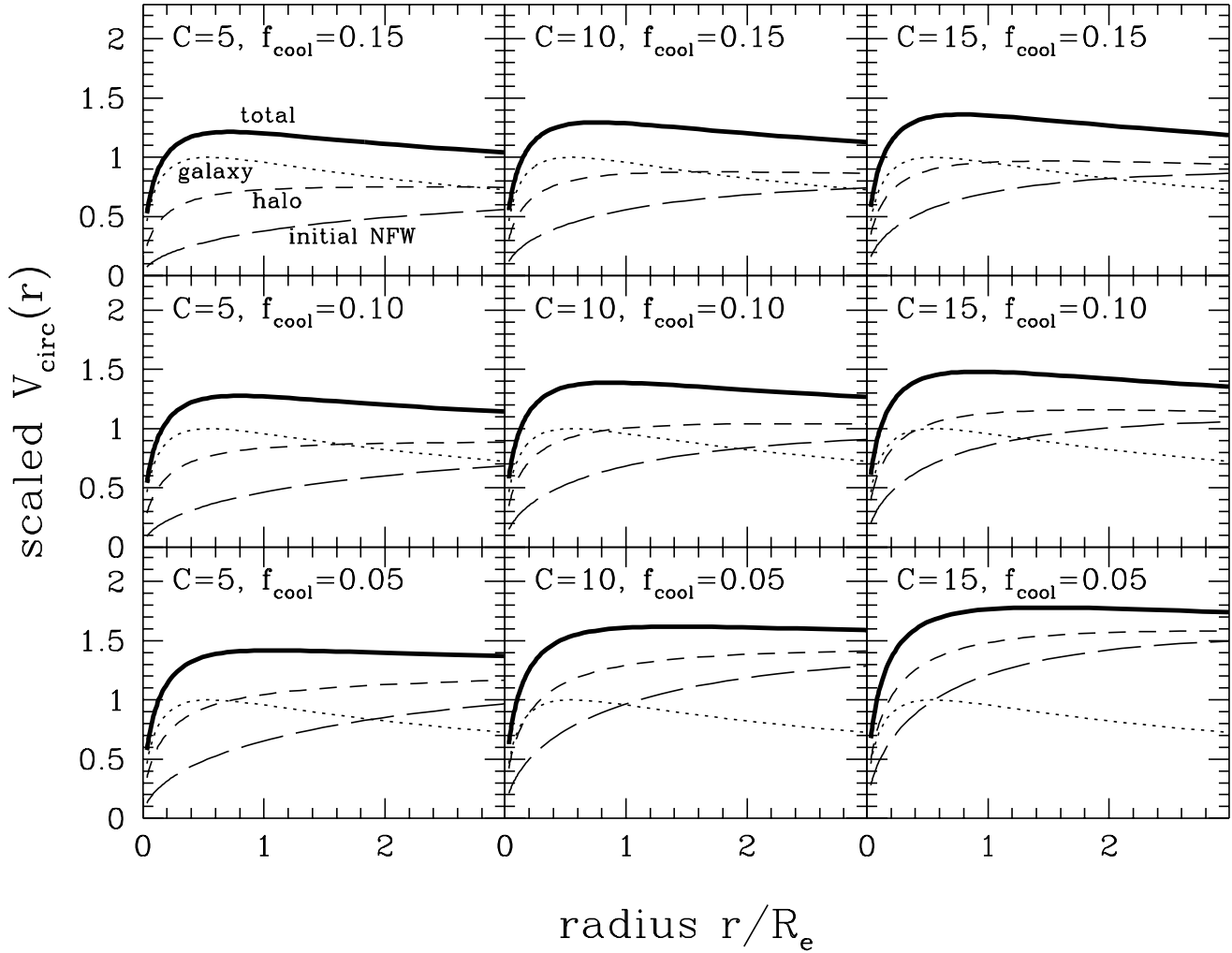}}
\caption{
Rotation curves for sample star+halo models with NFW halos. Each
panel has the specified values of the concentration $C$ of the
initial halo and the cooled mass fraction $\fcool$; all models have
$R_e/r_{200} = 0.03$. The solid curves show the total rotation
curves, while the dotted and dashed curves show the contributions
from the galaxy and halo, respectively. For comparison, the long
dashed curves show the rotation curves of the initial NFW halos
before adiabatic contraction. The velocities are scaled by the peak
velocity of the {\it galaxy\/} component.
}\label{fig:rotcrv1}
\end{figure*}

The dark matter halo models were derived from studies of
collisionless dark matter. They do not hold in the presence of
dissipative baryons, because as the baryons cool and condense
into a galaxy they modify the gravitational potential and thus
the dark matter distribution (e.g., Blumenthal et al.\ 1986;
Dubinski 1994). Fortunately, there is a simple analytic
prescription called adiabatic contraction for computing the
changes to the dark matter distribution; it seems to agree well
with gasdynamical simulations (e.g., Blumenthal et al.\ 1986;
Flores et al.\ 1993), even for merger scenarios thought to produce
elliptical galaxies (Gottbrath 2000). Appendix A gives an analytic
solution for adiabatic contraction of an arbitrary dark matter
halo by a Hernquist galaxy. Adiabatic contraction depends on the
mass ratio of the cooled galaxy component to the total virial
mass, $\fcool = M_{gal}/M_{tot}$, which is presumably no larger
than the global baryon fraction of the system, $\fbar =
M_{bar}/M_{tot} = \Omega_b/\Omega_M$. (There may be baryons that
remain hot and distributed throughout the halo, so $\fcool \le
\fbar$.) The virial mass and radius of the system factor out to
provide overall scalings, so the solution also depends on the
fraction $\fcool$ of the concentration $C$ of the initial halo,
and the effective radius of the galaxy (specifically $R_e/r_{200}$).

To illustrate the star+halo models, \reffig{rotcrv1} shows rotation
curves for various values of the parameters. For a fixed stellar
component, decreasing $\fcool$ increases the total mass of the halo
($M_{tot} = \fcool^{-1} \times M_{gal}$), which raises the rotation
curve. Increasing the concentration of the initial halo packs more
of the dark matter into the inner regions of the system, which also
raises the inner rotation curve. In other words, changing either
parameter affects the amount of mass contained within a few
effective radii of the galaxy. Lensing can distinguish between the
two parameters only if it is sensitive to the detailed shape of the
galaxy mass profile inside a few $R_e$.

\reffig{rotcrv1} offers two important qualitative results. First,
the galaxy and halo components can easily combine to produce a
rotation curve that is relatively flat from $\sim\!0.5 R_e$ to
several $R_e$. In other words, star+halo models can naturally
produce net mass distributions that are fairly close to $\rho
\propto r^{-2}$ throughout much of the galaxy. Second, comparing
the rotation curves of the halo before and after adiabatic
contraction illustrates that the modification by the baryons can
significantly increase the halo mass within a few $R_e$, especially
for less concentrated halos. \reffig{rotcrvB} also shows that
adiabatic contraction affects NFW profiles more dramatically than
Moore profiles, especially for large $\fcool$, which tends to
reduce the differences between NFW and Moore model galaxies.

\begin{figureone}
\centerline{\epsfxsize=3.5in \epsfbox{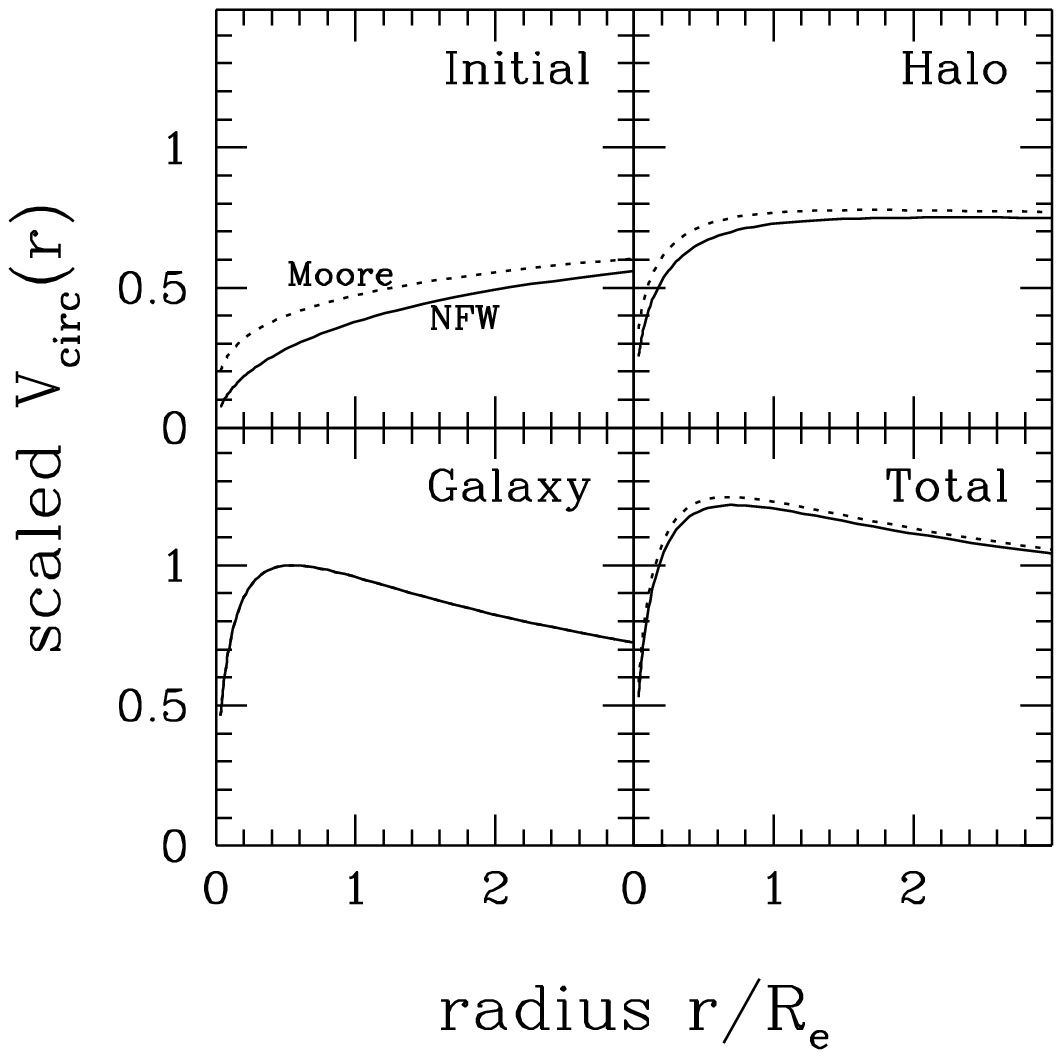}}
\caption{
A comparison of rotation curves for NFW and Moore models. The
four panels show the various components of the rotation curve.
The solid curves indicate NFW models and the dotted curves show
Moore models. Results are shown for $C=5$, $\fcool=0.15$, and
$R_e/r_{200} = 0.03$.
}\label{fig:rotcrvB}
\end{figureone}
\vspace{0.2cm}

\subsection{Normalizations}

The CNOC2 field galaxy redshift survey (Lin et al.\ 1999, 2001)
gives the luminosity function of early-type galaxies at redshifts
$0.12 < z < 0.55$. The luminosity function is parametrized as an
evolving Schechter (1976) function, and Lin et al.\ (2001) give
parameter values for two cosmologies: a high-density, flat universe
with matter density $\Omega_M=1$; and a low-density, flat universe
with matter density $\Omega_M=0.2$ and cosmological constant
$\Omega_\Lambda = 0.8$. I use a Hubble constant $H_0 = 50\,\Hunits$
for the $\Omega_M=1$ cosmology to mimic the Standard Cold Dark
Matter cosmology, and $H_0 = 65\,\Hunits$ for the $\Omega_M=0.2$
flat cosmology. I convert the luminosity of the stellar component
into a mass using population synthesis models by Bruzual \& Charlot
(1993), modeling early-type galaxies with an old coeval stellar
population.

I place a dark matter halo around each galaxy and use empirical
correlations to normalize the galaxies and halos. Bright early-type
galaxies are observed to populate a ``fundamental plane'' in the
space of surface brightness, effective radius, and velocity
dispersion, with very little scatter away from this plane (e.g.,
Djorgovski \& Davis 1987; Dressler et al.\ 1987). Projecting out
the velocity dispersion yields a correlation between luminosity
and effective radius that has somewhat larger scatter but is
easier to use. Schade, Barrientos \& L\'opez-Cruz (1997) find that
at $z=0$ the $M(B)$--$\log R_e$ relation is
\begin{equation} \label{eq:magRe1}
  M_{AB}(B) - 5 \log h = -3.33 \log(R_e/\hkpc) - 18.15 \pm 0.06\,,
\end{equation}
and the relation evolves with redshift in a way that is consistent
with the fading of stellar populations due to passive evolution.
Early-type dwarf galaxies, on the other hand, appear to form a
population that is disjoint from giant galaxies. Binggeli \&
Cameron (1991, 1993) demonstrate this effect in the Virgo cluster,
and a fit to their data yields
\begin{equation} \label{eq:magRe2}
  M_{AB}(B) = -13.1 \log(R_e/\kpc) - 13.7\,,
\end{equation}
although with significant scatter. The break between giant and
dwarf galaxies occurs somewhere around an absolute magnitude of
$-16$ or $-18$, but it is not sharp. The exact location of the
break has little effect on lens statistics because these low-mass
galaxies contribute little lensing optical depth.

Simulated dark matter halos do not all have the same profile.
Halos of a given mass have a range of concentrations; and
cluster-mass halos are systematically less concentrated than
galaxy-mass halos (e.g., Jing \& Suto 2000; Bullock et al.\ 2001).
Bullock et al.\ (2001) characterize the scatter by fitting NFW
profiles\footnote{Bullock et al.\ (2001) remark that it would
be possible to fit other profiles to halos but argue that the
\refeq{Cprob} captures the full range of halo properties seen in
their simulations.} to simulated halos and obtaining a set of
concentration parameters consistent with the log--normal
distribution
\begin{equation} \label{eq:Cprob}
  p(\log C | M, z) = {1 \over \sqrt{2\pi}\,\sigma_C}\,
    \exp\left\{- {(\log[C/C_{med}(M,z)])^2 \over 2\sigma_C^2}
    \right\} ,
\end{equation}
where $\sigma_C = 0.18$, and the median concentration varies
systematically with mass and redshift as $C_{med}(M,z) \propto
M^{-1/9} (1+z)^{-1}$.  The scatter in halo properties is
important for lensing because more concentrated halos are much
better lenses.  To include this effect, I use halos drawn
randomly from \refeq{Cprob}.  I normalize the distribution in
terms of the parameter $\CC$ defined to be the median concentration
of $10^{12}\,\hMsun$ halos at redshift $z=0$.  The value of $\CC$
is predicted by simulations (see \S 4.2), but I take it to be a
free model parameter.

\section{Lensing Methods}

The adiabatic contraction solution gives the mass profile $M(r)$ of
the final system. The system's projected surface density $\kappa(R) =
\Sigma(R)/\Sigma_{cr}$ and lensing deflection $\phi_{R}$ can then
be written as (see Keeton 2001)
\begin{eqnarray}
  \kappa(R) &=& {\kappa_{200} \over 2} \int_R^\infty dr\,
    {m'(r) \over r \sqrt{r^2-R^2}}\ , \label{eq:kap1} \\
  &=& {\kappa_{200} \over 2R} \int_0^1 dy\, {1 \over 1+y^2}\, \times
    \label{eq:kap2} \\
  &&\ \left[ m'\left(R\sqrt{1+y^2}\right) 
    + m'\left(R\sqrt{1+y^{-2}}\right) \right], \nonumber \\
  \phi_{R}(R)
  &=& \kappa_{200}\,r_{200}\,R \int_R^\infty dr\,
    {m(r) \over r^2 \sqrt{r^2-R^2}}\ , \label{eq:def1} \\
  &=& {\kappa_{200}\,r_{200} \over R} \int_0^1 dy\,
    {1 \over (1+y^2)^{3/2}}\, \times \label{eq:def2} \\
  &&\ \left[ m\left(R\sqrt{1+y^2}\right)
    + y\,m\left(R\sqrt{1+y^{-2}}\right) \right], \nonumber
\end{eqnarray}
where the radii are written in units of $r_{200}$, the mass
$m(r)=M(r)/M_{200}$ is written in units of the total mass inside
$r_{200}$, and $m'(r) = dm(r)/dr$. \refEqs{kap2}{def2} represent
variable transformations that give the integrals a finite range,
which is useful for numerical integration. The strength of the
system as a gravitational lens is measured by the dimensionless
parameter
\begin{eqnarray}
  \kappa_{200} &=& {M_{200} \over \pi\,r_{200}^2\,\Sigma_{cr}}
    \label{eq:k200} \\
  &=& 0.00467 \left[{M_{200} \over 10^{10}\,\hMsun}\right]^{1/3}\,
    \left[{H(z) \over H_0}\right]^{4/3}\,{D_l\,D_{ls} \over r_H\,D_s}\ ,
    \nonumber
\end{eqnarray}
where $r_H = c/H_0$ is the Hubble distance. This parameter is the
mean projected surface density of the system in units of the
critical density for lensing. In general, $\kappa_{200}$ is
considerably less than unity because most halos can act as strong
gravitational lenses only in a high-density region near the core,
not all the way out to the virial radius.

The images corresponding to a given source are found by solving the
lens equation,
\begin{equation}
  u = R - \phi_{R}(R)\,,
\end{equation}
where $u$ is the angular position of the source relative to the
lens (see Schneider et al.\ 1992 for a full discussion). The
magnification of an image at position $R$ is
\begin{equation}
  \mu(R) = (1-\phi_{R}/R)^{-1} (1-\phi_{RR})^{-1} .
\end{equation}
Here $\phi_{RR} = d(\phi_R)/dR$, which can be computed efficiently
using the identity $R^{-1}\,\phi_{R} + \phi_{RR} = 2\kappa$. In
general, a spherical lens has two radii at which the magnification
is infinite. These radii correspond to ``critical curves'' in the
image plane, which map to ``caustics'' in the source plane. The
outer or tangential critical curve lies at the Einstein ring radius
$r_E$ of the lens; a source directly behind the lens produces a
ring image with radius $r_E$. The inner or radial critical curve
lies at a small radius $r_{cr}$. The source position corresponding
to an image at $r_{cr}$, which I label $u_{out}$, marks the boundary
of the region where lensing yields multiple images. (The equations
for $r_{cr}$ and $u_{out}$ are given in Appendix B.) A source with
$u<u_{out}$ has three images, one outside $r_E$, one between $r_E$
and $r_{cr}$, and one inside $r_{cr}$; the innermost image is usually
demagnified and undetected (see \S 5). A source with $u>u_{out}$ has
a single image, which is outside $r_E$.

Computing the statistics of gravitational lenses requires summing
over populations of lenses and sources, and accounting for
``magnification bias,'' or the fact that a flux-limited survey may
include lenses where the source is intrinsically fainter than the
flux limit, but lensing magnification brings the object into the
sample (e.g., Turner 1980; Turner, Ostriker \& Gott 1984). The
number of lenses with a total flux greater than $S$ expected to be
found in a survey with lensing selection functions described by
$\F$ is
\begin{eqnarray}
  N_\lens(>\!S) &=& {1 \over 4\pi} \int dz_s \int_{\obs}^{\src} dV
    \int dM\,{dn \over dM} \label{eq:Nlens} \\
  && \quad\times \int d(\log C)\,p(\log C | M, z_l) \nonumber\\
  && \quad\times \int_{0}^{u_{out}} du\,2\pi u\, \F(u)\,
    {dN_\src(>\!S/\mu) \over dz_s}\ , \nonumber
\end{eqnarray}
where $z_s$ is the source redshift, $dV$ is the comoving volume
element (see, e.g., Carroll, Press \& Turner 1992), and $dn/dM$
is the mass function of halos that can serve as lenses.  The
integral over $C$ incorporates the scatter in halo properties
defined in \refeq{Cprob}.  The factor ${\cal F}(u)$ indicates
whether a lens associated with a source at $u$ would be detected
given the selection functions. The volume, mass, and concentration
integrals sum over the population of possible lens galaxies. The
$z_s$ integral allows for a distribution of source redshifts,
where $[dN_\src(>\!S)/dz_s]\,dz_s$ is the number of sources
brighter than flux $S$ that lie in the redshift range $z_s$ to
$z_s+dz_s$.  Finally, the distribution of image separations is
found by computing $dN_\lens/d\theta$, and mean quantities are
found by averaging over the predicted lens population.

\section{The Number and Sizes of Lenses}

In this section, the global properties of the star+halo models are
evaluated using two quantities from lens statistics: the number of
lenses, or more specifically the fraction of sources that are
multiply imaged; and the distribution of lensed image
separations.\footnote{A third interesting quantity is the ratio of
four-image lenses to two-image lenses, which can be used to
constrain the angular shape of lensing mass distributions (e.g.,
Kochanek 1996; Rusin \& Tegmark 2001). This test requires
non-spherical lens models, and it is not very sensitive to the mass
profile of lensing halos.} Section 4.1 reviews the data. Section
4.2 presents results for a fiducial set of models, while Section
4.3 considers systematic effects including the source redshift
distribution, the galaxy formation redshift, and the density
profile.

\subsection{Data}

More than 50 galaxy-mass lenses are known, and their properties
have been compiled by the CfA/Arizona Space Telescope Lens Survey
(CASTLES; see Kochanek et al.\ 2001). This sample includes lenses
from a variety of surveys as well as serendipitous discoveries;
thus, the parent (or source) population is unknown, and the CASTLES
sample cannot be used to test the number of lenses. By contrast,
the distribution of image separations in the sample probably can be
used, because it is insensitive to the size of the source
population.

The largest homogeneous statistical survey for lenses is the Cosmic
Lens All-Sky Survey (CLASS; Helbig 2000; Browne 2001). The sample
comprises $10,499$ flat-spectrum radio sources with flux $S>30$ mJy
at 5 GHz, and the flux distribution can be described as a power law
$dN_\src/dS \propto S^\nu$ with $\nu \approx -2.1$ (see Rusin \&
Tegmark 2001). The survey includes 18 lenses, all of which have
image separations $\theta < 3\arcsec$, and the survey is believed
to be complete at image separations $0.3\arcsec < \theta <
15\arcsec$ (Helbig 2000; Phillips et al.\ 2001). Because this paper
focuses on lensing by elliptical galaxies, I omit two CLASS lenses
that are known to be produced by spiral galaxies (B~0218+357 and
B~1600+434). For the CLASS lenses where the lens galaxy type is not
known, I assume an elliptical galaxy because most lens galaxies are
ellipticals (e.g., Kochanek et al.\ 2000) and because this is the
conservative approach (as shown below). In the CLASS survey both
the source and lens populations are known, so the sample can be used
to test both the number of lenses and the distribution of image
separations.

\begin{figure*}[t]
\centerline{\epsfxsize=7.0in \epsfbox{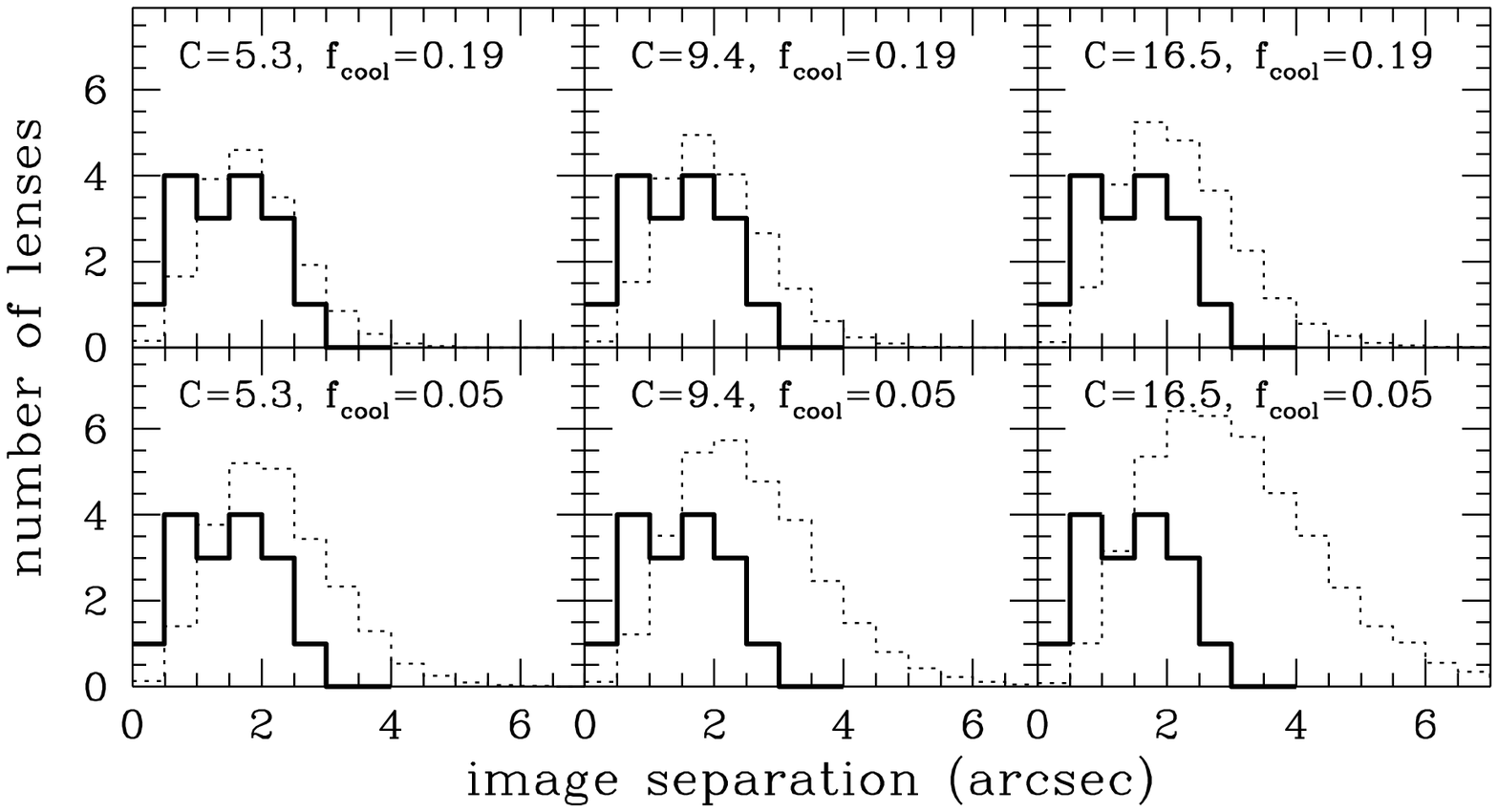}}
\caption{
Image separation histograms for the CLASS data (solid lines) and
for sample models (dotted lines). Model results are shown for the
fiducial models in an $\Omega_M=0.2$ flat cosmology. The model
parameters are indicated in each panel.
}\label{fig:hist}
\end{figure*}

The number of lenses can be tested (the ``N test'') by using Poisson
statistics to compare the CLASS sample with predictions from the
models.  The image separations can be tested (the ``$\theta$ test'')
by using a Kolmogorov--Smirnov, or K--S, test (e.g., Press et al.\
1992) to compare the observed and predicted distributions of image
separations.  The CLASS and CASTLES samples can both be used for
the $\theta$ test, although the tests are not independent because
the CASTLES sample contains the entire CLASS sample. The two samples
are consistent in the sense that a K--S test does not reveal a
significant difference between their separation distributions. The
CASTLES sample is larger and thus less sensitive to statistical
peculiarities (such as the lack of CLASS lenses with
$\theta > 3\arcsec$, perhaps). The CLASS sample, however, has better
information about source fluxes and redshifts. In an attempt to
compromise between the two samples, I perform the $\theta$ test with
both samples and conservatively adopt the {\it weaker\/} of the two
results.

The largest uncertainty in the models arises from the source
redshift distribution, which is not known for the full CLASS sample.
Marlow et al.\ (2000) report redshifts for a small subsample of 27
sources from the CLASS sample. They find a mean redshift of
$\avg{z}=1.27$, which is comparable to that found in other radio
surveys at comparable fluxes (Drinkwater et al.\ 1997; Henstock et
al.\ 1997; Falco, Kochanek \& Mu\~noz 1998). They also find evidence
for a difference between the galaxy and quasar populations in the
sample, with $\avg{z_{gal}} = 0.18$ for 8 galaxies and
$\avg{z_{QSO}} = 1.72$ for 19 quasars. It is not clear at this
point whether the subsample fairly represents the full sample.
To examine possible systematic effects, I consider a set of models
with all sources placed at the mean redshift of the subsample,
and an alternate set of models with source redshifts distributed
according to the subsample.

\subsection{Basic results}

Consider a fiducial set of models in which the halos before adiabatic
contraction are modeled with NFW profiles, the galaxies have old
stellar populations that formed at redshift $z_f=5$, and all the
sources are placed at the mean redshift of the CLASS spectroscopic
subsample, $z_s=1.25$.  \reffig{hist} compares model predictions with
the data from the CLASS sample.  As the median\footnote{Recall that
the calculation explicitly includes scatter in the halo properties
(see eq.~\ref{eq:Cprob}), so the models are characterized by the
median concentration.} concentration $\CC$ increases or the cooled
mass fraction $\fcool$ decreases, the number of lenses increases and
the distribution of image separations shifts to higher values.
Physically, increasing $\CC$ or decreasing $\fcool$ raises the
amount of dark matter in the inner parts of halos (see \S 2.1),
leading directly to more and larger lenses.

\begin{figure*}[t]
\centerline{\epsfxsize=7.0in \epsfbox{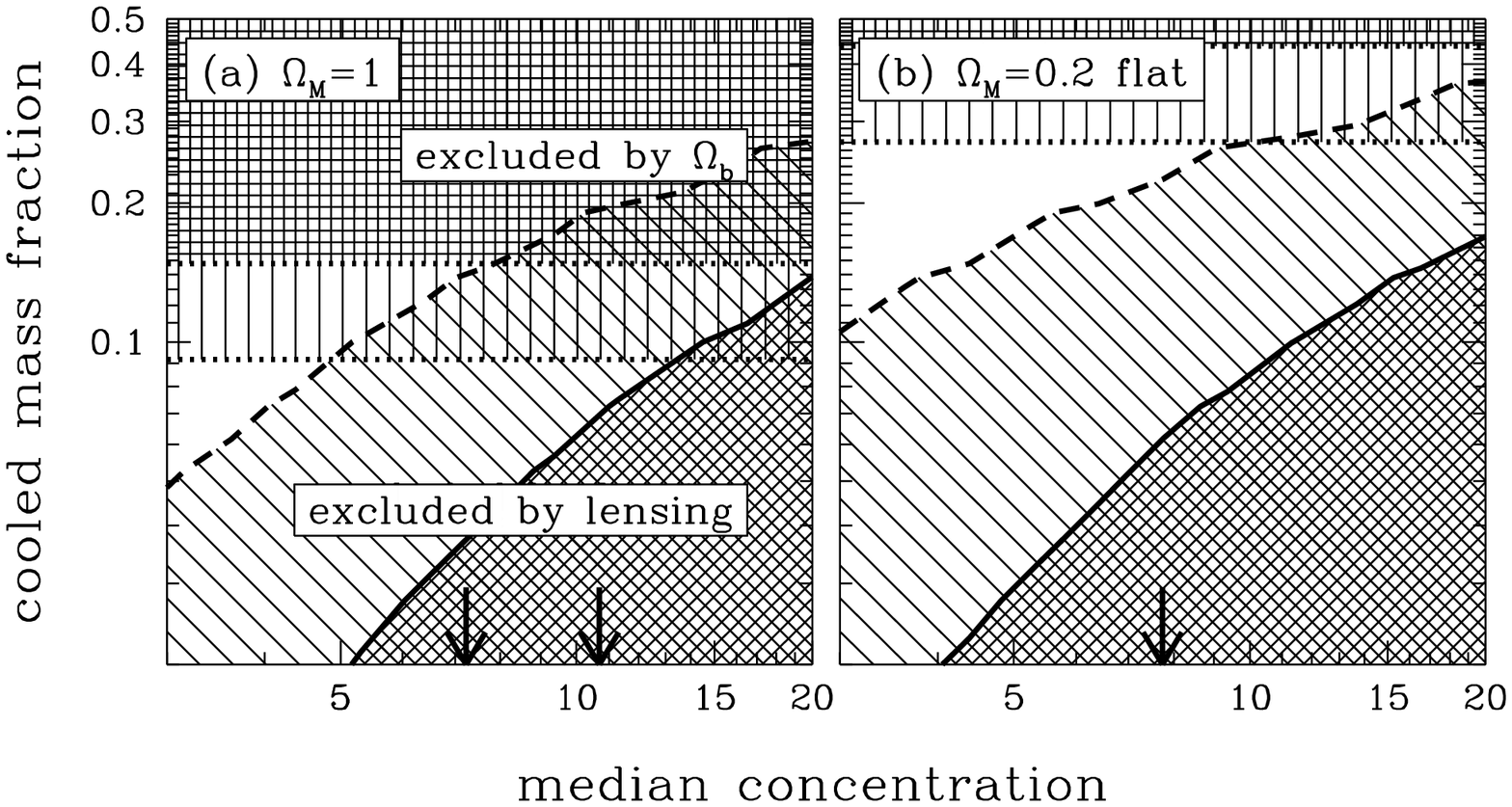}}
\caption{
Confidence regions in the $(\CC,\fcool)$ plane for the fiducial
models. The shaded regions below the diagonal curves are excluded
at 95\% confidence by the lens data; the lower and upper curves
correspond to the $N$ and $\theta$ tests, respectively. The shaded
regions above the horizontal lines are excluded by measurements of
the cosmic baryon density $\Omega_b$. The lower curve corresponds
to $\Omega_b h^2 = 0.019\pm0.0024$ from measurements of deuterium
(Tytler et al.\ 2000), and the upper curve corresponds to $\Omega_b
h^2 < 0.037$ (95\% confidence) from the cosmic microwave background
(Tegmark et al.\ 2001). The arrows on the $x$-axes indicate
concentrations predicted by CDM simulations (see text). Results are
shown for two cosmologies.
}\label{fig:clev1}
\end{figure*}

Comparing the models to the data using the $N$ and $\theta$
statistical tests yields confidence limits on the model parameters,
as shown in \reffig{clev1}. There is a band in the upper left of
the $(\CC,\fcool)$ plane where the models are consistent with both
the number of lenses and the distribution of image separations.
Moving to larger $\CC$ or smaller $\fcool$ increases the number
and sizes of predicted lenses. In the hatched region, the predicted
lenses are generally too big, and the models are ruled out (at
95\% confidence) by the $\theta$ test. In the cross-hatched region,
the models are further excluded because they predict too many
lenses (the $N$ test).\footnote{At small $\CC$ and large $\fcool$,
the models predict {\it too few\/} lenses that are {\it too small\/}
compared with the data. These constraints apply beyond the upper
left corner of the $(\CC,\fcool)$ plane in \reffig{clev1}.} The
contours from the $N$ test are based on the assumption that 16
CLASS lenses are produced by elliptical galaxies. If the number
of CLASS lenses with elliptical galaxies turns out to be smaller
than 16, the $N$ contours would move further up and to the left,
strengthening the constraints from lensing.

There is little difference between the lensing constraints in the
two cosmologies shown in \reffig{clev1}, which seems surprising
because it is traditionally argued that lens statistics are quite
sensitive to a cosmological constant $\Lambda$ (e.g., Turner 1990;
Kochanek 1996). The traditional argument is based on models where
the lens galaxy population is obtained by taking the local comoving
number density of galaxies and assuming that it holds out to
redshift $z \sim 1$ in all cosmologies; in this case, the number of
lenses is very sensitive to the volume of the universe to $z \sim 1$,
and hence to $\Lambda$. By contrast, my models are based on counts
of galaxies at $z \sim 0.5$ from the CNOC2 field galaxy redshift
survey (Lin et al.\ 1999, 2001).  In these models, the volume factor
required to convert from number counts to number density (or
luminosity function) essentially cancels the volume factor that
appears in the lensing analysis; the number of lenses is roughly
proportional to the number counts of galaxies, and is not very
sensitive to $\Lambda$. In other words, using models normalized by
number counts of galaxies at $z \sim 0.5$ makes lens statistics only
weakly sensitive to cosmology.

CDM simulations make specific predictions about the concentration:
in a cosmology with $\Omega_M=1$, $\CC \simeq 11$ for standard CDM,
or $\CC \simeq 7$ for tilted CDM where the power spectrum has shape
parameter $\Gamma=0.2$; and in an $\Omega_M=0.2$ flat cosmology,
$\CC \simeq 8$ (Navarro et al.\ 1997). These values are indicated
by arrows on the $x$-axes in \reffig{clev1}.  Cosmic baryon censuses
give limits on $\fcool$.  Because $\fcool$ gives the fraction of
a system's mass that has cooled into the baryonic galaxy, it is a
lower limit on the baryonic content of the system and hence should
not exceed the cosmic baryon fraction, $\Omega_b/\Omega_M$ (e.g.,
White et al.\ 1993). The upper limits on $\fcool$ derived from
measurements of $\Omega_b$ using the cosmic microwave background
(e.g., Tegmark, Zaldarriaga \& Hamilton 2001) and the
deuterium/hydrogen ratio and big bang nucleosynthesis (e.g., Tytler
et al.\ 2000) are indicated by horizontal lines in \reffig{clev1}.
Some elliptical galaxies contain hot, X-ray emitting gas that is
probably primordial gas that never cooled; the cool stellar
component may contain as little as half of the baryons (e.g.,
Brighenti \& Mathews 1998). The presence of hot gas would reduce
the upper limit on $\fcool$ to something below $\Omega_b/\Omega_M$;
but because the actual amount of gas and its presence across the
galaxy population are not well understood, I focus on the
conservative upper limit from $\Omega_b$.

The lens data reject models where concentrated, massive dark matter
halos make elliptical galaxies overly efficient lenses --- a large
portion of the $(\CC,\fcool)$ plane.  At the concentrations found
in CDM simulations, lensing requires baryon fractions that are
incompatible with a high-density universe. The lens data are formally
compatible with a low-density universe, but only in a narrow corner
of parameter space where galaxy-mass halos must be very efficient
at cooling their baryons. The general conclusion, then, is that
galaxies constructed from CDM mass distributions are too concentrated
to agree with lens statistics, especially in a high-density CDM
cosmology.

There is clearly a degeneracy between $\CC$ and $\fcool$ in
\reffig{clev1}, which is not surprising because both parameters
affect the central mass that determines the lensing properties. It
is therefore interesting to define an integral quantity,
\begin{equation} \label{eq:Mdef}
  \M(r) \equiv \left\langle {M_{halo}(r) \over M_{gal}(r)} \right\rangle ,
\end{equation}
which is the ratio of halo mass to galaxy mass inside some radius
$r$, where the average is over the lens population. Note that $\M$
is defined using the mass in spheres. The mass ratio allows
model-independent statements about the dark matter distribution
in early-type galaxies, which are summarized in Table 1. In
the fiducial models, dark matter can account for up to 29--33\% of
the mass inside $R_e$ (95\% confidence upper limit), and up to 35--40\%
of the total mass inside $2 R_e$. In other words, dark matter can
contribute a moderate fraction of the mass in the inner regions of
elliptical galaxies, but it is not the dominant mass component at
small radii. These interesting upper limits result from the
distribution of image separations; observed lenses are too small to
be consistent with larger dark matter contributions. The lensing
limits are similar to but stronger than those derived from a dynamical
analysis of the nearby elliptical galaxy NGC 2434 (Rix et al.\ 1997).
They are consistent with the lower limits on dark matter in ellipticals
derived from the relationship between X-ray temperature and stellar
velocity dispersion (Loewenstein \& White 1999).

\vspace{0.1in}
\begin{tableone}
\begin{center}
\begin{tabular}{cccc}
\multicolumn{4}{c}{\sc Table 1} \\
\multicolumn{4}{c}{\sc Halo/Galaxy Mass Ratio} \\
\tableline
\tableline
 & Radius & $\Omega_M=1$ & $\Omega_M=0.2$ flat \\
\tableline
 Case 1 & $  R_e$ & $\M < 0.50$ & $\M < 0.41$ \\
        & $2 R_e$ & $\M < 0.66$ & $\M < 0.55$ \\
 Case 2 & $  R_e$ & $\M < 0.43$ & $\M < 0.27$ \\
        & $2 R_e$ & $\M < 0.57$ & $\M < 0.35$ \\
 Case 3 & $  R_e$ & $\M < 0.50$ & $\M < 0.44$ \\
        & $2 R_e$ & $\M < 0.68$ & $\M < 0.60$ \\
 Case 4 & $  R_e$ & $\M < 0.50$ & $\M < 0.41$ \\
        & $2 R_e$ & $\M < 0.65$ & $\M < 0.52$ \\
\tableline
\end{tabular}
\end{center}
Note. ---
95\% confidence upper limits on the halo/galaxy mass ratio,
defined in \refeq{Mdef}, computed at two radii for two
cosmologies. The four different cases are defined in the text.
\end{tableone}

\subsection{Systematic effects}

\begin{figure*}[t]
\centerline{\epsfxsize=7.0in \epsfbox{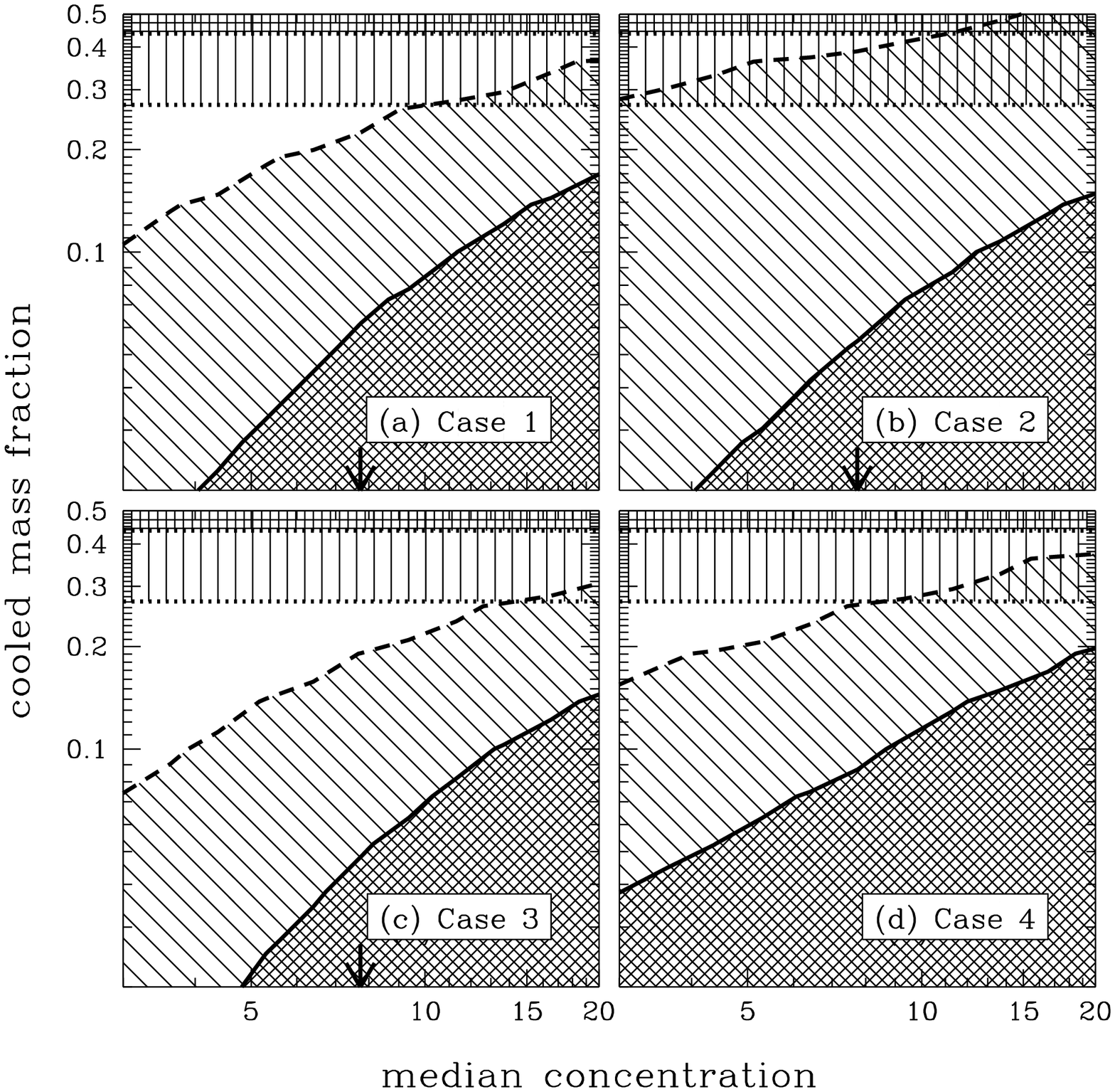}}
\caption{
Confidence regions in the $(\CC,\fcool)$ plane, shown for the four
cases defined in the text, in an $\Omega_M=0.2$ flat cosmology.
Panel (a) is the same as \reffig{clev1}b but is included here for
completeness.
}\label{fig:clevB}
\end{figure*}

There are three systematic effects that may be important for the
models. \reffig{clevB} shows how the results from the fiducial
models (case 1, \reffig{clevB}a) are changed by each effect, and
Table 1 gives the updated constraints on the halo/galaxy mass
ratio $\M$. In case 2 (\reffig{clevB}b), the fixed source redshift
is replaced by a redshift distribution to match the CLASS
spectroscopic subsample (Marlow et al.\ 2000).  The new models
predict larger image separations, so the region excluded by the
$\theta$ test stretches up and to the left. While the entire
$(\CC,\fcool)$ plane is now formally excluded either by lensing
or by $\Omega_b$, it is not clear how strongly to interpret this
result, because the CLASS spectroscopic subsample may not fairly
represent the full CLASS sample.  The strength of the conclusions
will ultimately be limited by the extent to which the redshift
distribution of the full CLASS sample can be determined.
Nevertheless, it is important to discover that the redshift
distribution may actually worsen the discrepancy between the data
and the models.

In case 3 (\reffig{clevB}c), the redshift at which the stellar
populations formed is reduced from $z_f=5$ to $z_f=3$. The younger
stellar populations have smaller mass-to-light ratios, so the
galaxies are poorer lenses (because the luminosity function is held
fixed). Thus, the models predict fewer and smaller lenses, and the
regions excluded by lensing move down and to the right in the
$(\CC,\fcool)$ plane. Nevertheless, the changes in the lensing
constraints are small; lens statistics are not very sensitive to
the galaxy formation redshift, provided that the stellar
populations of elliptical galaxies are old. The Fundamental Plane
of elliptical galaxies in rich clusters out to $z=0.83$ (e.g., van
Dokkum et al.\ 1998) and of elliptical lens galaxies in low-density
environments out to $z \sim 1$ (Kochanek et al.\ 2000) indeed
implies old stellar populations, $z_f \gtrsim 2$.

Finally, in case 4 (\reffig{clevB}d), the initial NFW profiles are
replaced by steeper Moore profiles. Moore halos have more mass in
the central regions than NFW halos, even for a fixed concentration
parameter, and thus yield better lenses. Hence, the models predict
more and larger lenses, and the excluded regions move up and to the
left in the $(\CC,\fcool)$ plane. The change is not very dramatic,
however, because of the effects of adiabatic contraction. Moore
halos, which are denser than NFW halos to begin with, experience a
smaller density enhancement under adiabatic contraction (see
\reffig{rotcrvB}). In other words, adiabatic contraction tends to
erase some of the differences between NFW and Moore models.

These results suggest that systematic effects do not weaken the
discrepancy between models and data, and may even strengthen it.
CDM star+halo models are at best marginally consistent with the
statistics of strong lenses, and may be quite inconsistent
depending on the distribution of source redshifts in the full CLASS
sample. As a relatively model-independent conclusion, the lensed
image separations imply that dark matter can contribute no more
than about 33\% of the total mass inside $R_e$, or about 40\% of
the mass inside $2 R_e$ (95\% confidence; see Table 1). CDM
halos appear to be too concentrated to agree comfortably with this
constraint.

\section{Odd images}

In this section, the very inner regions of star+halo models are
evaluated with lensing. The fact that most lenses do not show the
expected central or ``odd'' images places strong lower limits on
the central densities of galaxies. Section 5.1 reviews the data,
Section 5.2 presents results, and Section 5.3 offers a discussion.

\subsection{Data}

\begin{figure*}[t]
\centerline{\epsfxsize=7.0in \epsfbox{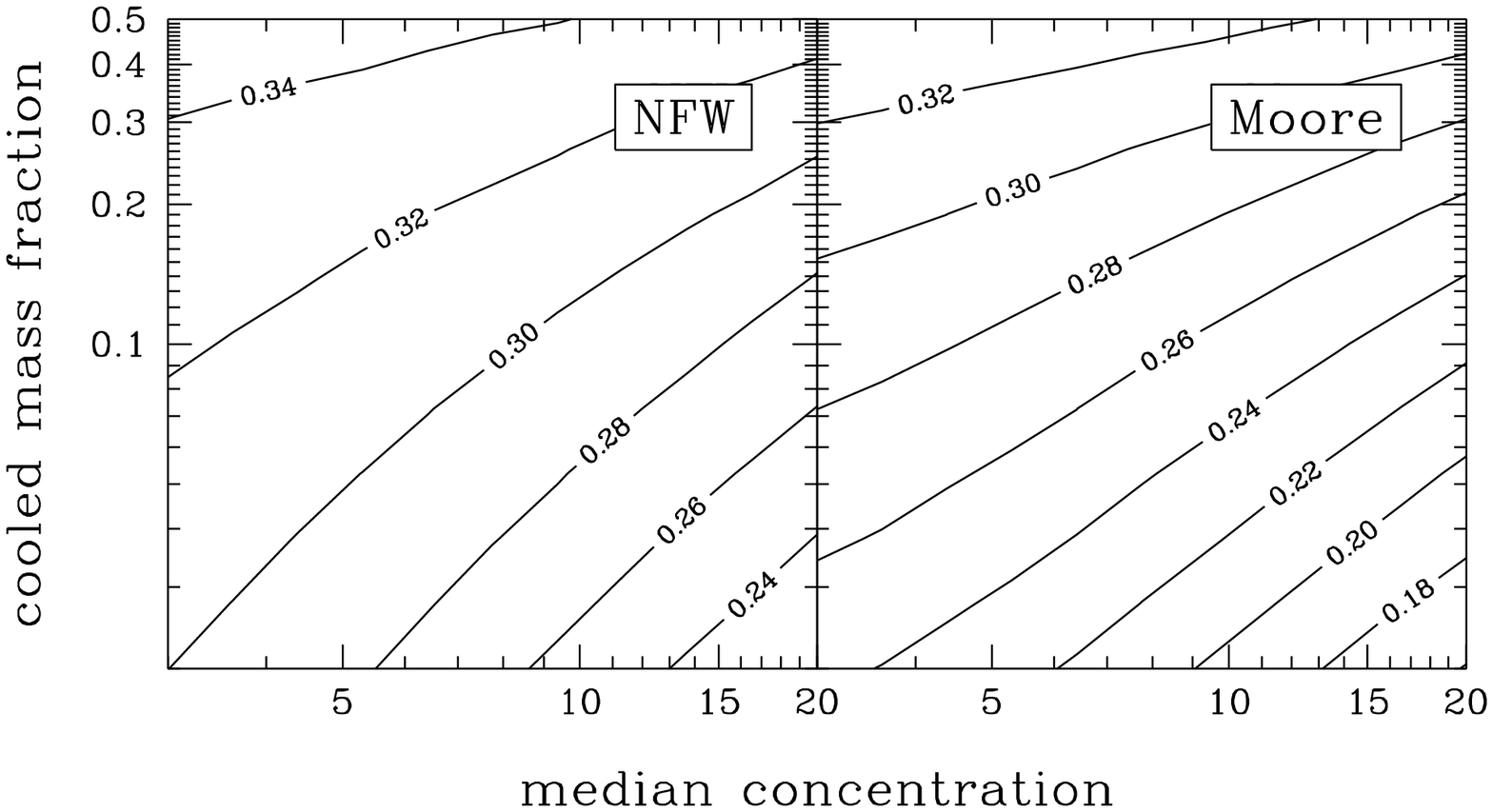}}
\caption{
Contours of the fraction of (two-image) lenses where the odd image
is brighter than 1\% of the brightest image ($\fodd \ge 0.01$).
Results are shown for an $\Omega_M=0.2$ flat cosmology. The left
panel shows models with NFW halos (case 1), and the right panel
show models with Moore halos (case 4).
}\label{fig:core}
\end{figure*}

It can be proved mathematically that a single thin lens with a
smooth (i.e., non-singular) projected mass density and a finite
mass always produces an {\it odd\/} number of images (Burke 1981;
also see Schneider et al.\ 1992). In other words, lenses are
generally expected to have three or five images, but are usually
observed to have two or four. The apparent paradox is resolved
by noting that for lenses with high central densities, one of the
images is close to the center of the lens and
demagnified.\footnote{Alternatively, if the projected mass density
is singular the odd image theorem formally breaks down.  Odd images
still appear, though, provided the central density cusp is shallower
than $\rho \propto r^{-2}$ (see Appendix B).}
If the central density is high enough, the central image may be
highly demagnified and therefore very difficult to detect. For
example, Appendix B shows that for an isothermal sphere with a
small core radius, or for a power law density $\rho \propto
r^{-\alpha}$ with $\alpha \approx 2$, the mean magnification of
central or ``odd'' images can be quite small. Every two- or
four-image lens may therefore be a three- or five-image system
where one of the images remains undetected.

Odd images are expected to be rare in optical observations, because
they would be swamped by light from the lens galaxies. The only
lens with an odd number of optical images is APM~08279+5255 (Ibata
et al.\ 1999); the third image is either a standard odd image, in
which case it requires a shallow density cusp $\alpha \lesssim 0.4$
for $\rho \propto r^{-\alpha}$ (Mu\~noz et al.\ 2001), or else it
represents a special image configuration produced by an edge-on
disk (Keeton \& Kochanek 1998). Radio observations should be much
more sensitive to odd images, because the lens galaxies (as opposed
to the sources) are rarely radio loud. The only candidate odd image
detected in the radio is in MG~1131+0456 (Chen \& Hewitt 1993),
although the possibility that the lens galaxy is radio loud cannot
be ruled out in this case.

The CLASS lens sample offers high-resolution and high-dynamic range
(noise level $\sim$50 $\mu$Jy beam$^{-1}$) radio maps of lenses with
compact radio sources, and thus should be quite sensitive to odd
images. Consequently, the fact that no CLASS lens shows an odd image
leads to strong upper limits on how bright the odd images can be.
Rusin \& Ma (2001) tabulate the upper limits on six two-image CLASS
lenses.  To factor out the unknown brightness of the source, they
quote upper limits in terms of the flux ratio $\fodd$ defined to be
the flux of the odd image relative to the flux of the brightest image.
The 5$\sigma$ upper limits range from $\fodd < 0.0083$ for B~0739+366
(Marlow et al.\ 2001) to $\fodd < 0.00049$ for B~0218+357 (Biggs et
al.\ 1999). Norbury et al.\ (2001) give limits on odd images for
CLASS lenses with more than two images. When quantifying odd images
relative to other lensed images, four-image lenses are much more
sensitive to asymmetry in the lens galaxy than two-image lenses.
Hence, I restrict attention to the two-image CLASS lenses where
spherical models are sufficient for interpreting odd images.

\subsection{Results}

\begin{figure*}[t]
\centerline{\epsfxsize=7.0in \epsfbox{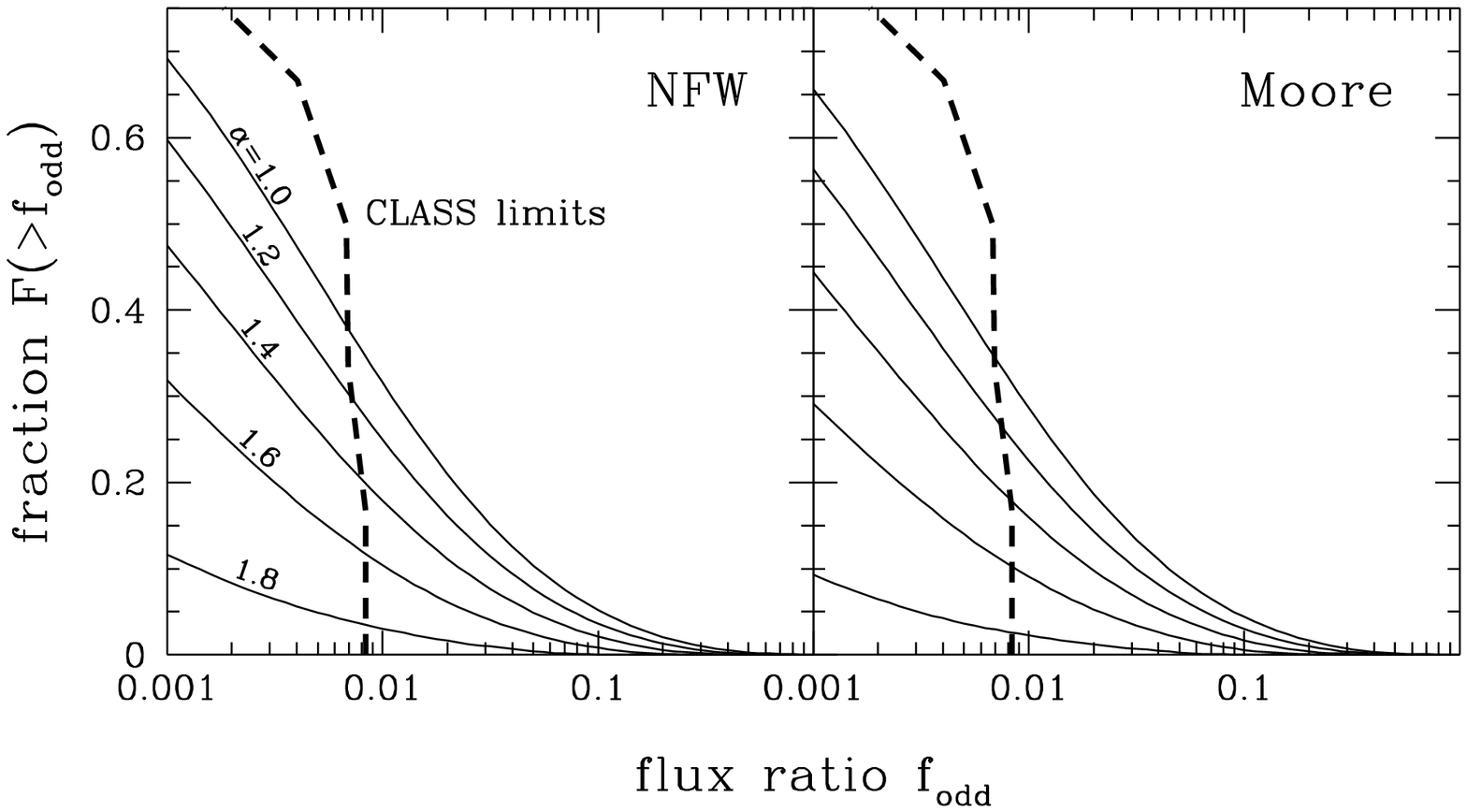}}
\caption{
The cumulative fraction of (two-image) lenses where the ratio of
the odd image to the brightest image is greater than $\fodd$.
Results are shown for models with $\CC=7.7$ and $\fcool=0.19$,
in an $\Omega_M=0.2$ flat cosmology. The galaxy components are
modeled as $\rho \propto r^{-\alpha} (r_s+r)^{\alpha-4}$, and
each curve shows results for a particular value of $\alpha$;
the fiducial Hernquist model corresponds to $\alpha=1.0$. The
initial mass distribution is modeled with an NFW (left panel)
or Moore (right panel) profile. The heavy dashed curves show
the {\it upper\/} limits derived from six two-image CLASS
lenses (Rusin \& Ma 2001).
}\label{fig:cusp}
\end{figure*}

\begin{figure*}[t]
\centerline{\epsfxsize=7.0in \epsfbox{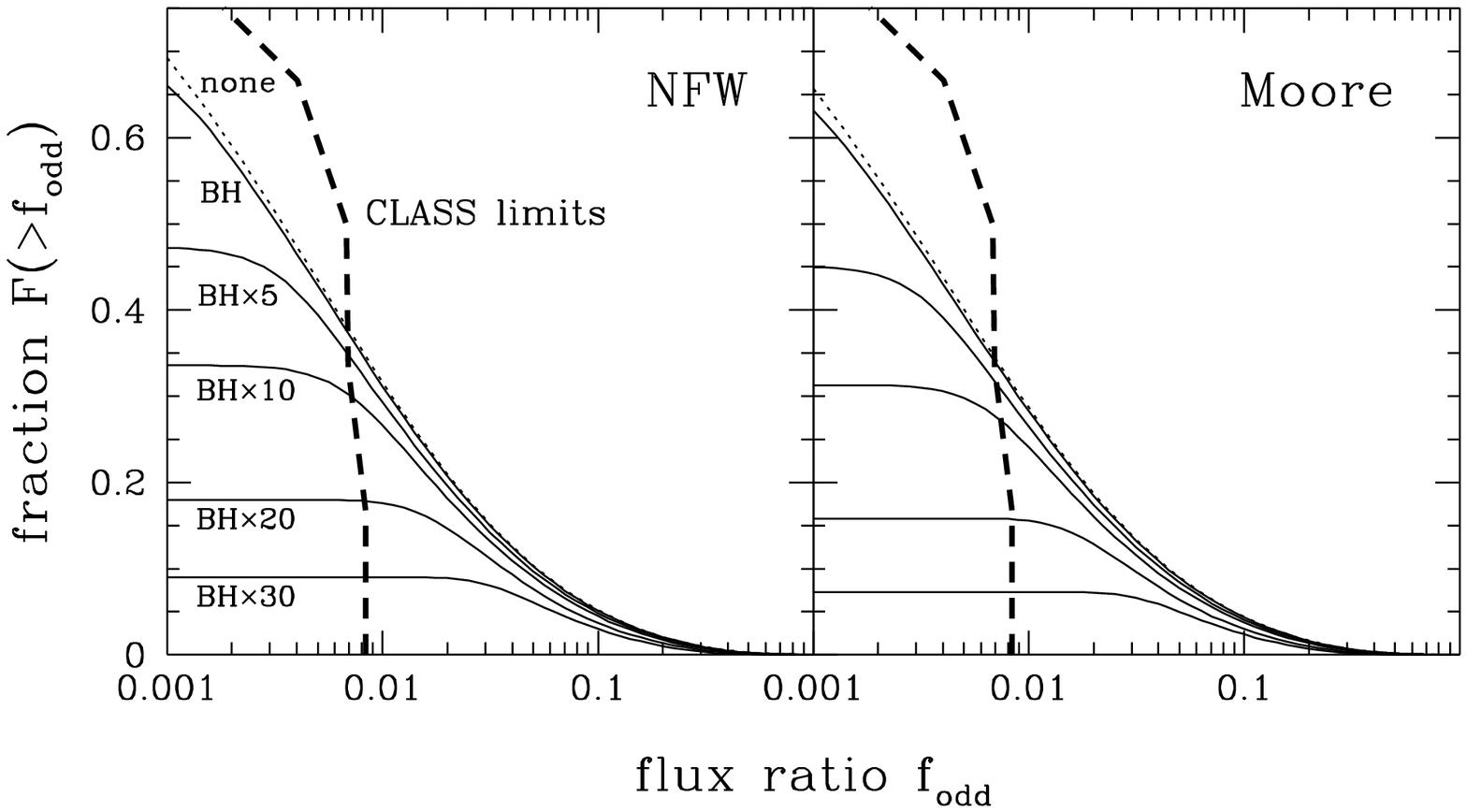}}
\caption{
Similar to \reffig{cusp}, but showing the effects of supermassive
black holes. The dotted curves show results for star+halo models
without black holes, for NFW (left panel) and Moore (right panel)
models. The solid curves show results when central black holes are
added. For the curves labeled ``BH,'' the black hole masses are
normalized by the empirical correlation between black hole mass
and galaxy velocity dispersion (Ferrarese \& Merritt 2000; Merritt
\& Ferrarese 2001). In the curves labeled ``BH$\times N$,'' the
black holes are made systematically more massive by the factor $N$.
(All galaxies have $\alpha=1.0$.) The heavy dashed curves again
show the upper limits from six CLASS.
}\label{fig:corebh}
\end{figure*}

For power law models with $\rho \propto r^{-1.5}$ the mean
magnification of odd images is unity, so odd images are not highly
demagnified (see eq.~\ref{eq:modd4} in Appendix B). This simple
prediction does not strictly apply to star+halo models, because
even in the initial halos the broken power laws affect lensing
via projection, and adiabatic contraction increases the central
density. However, it does suggest that the odd images in star+halo
models are worth investigating. The best way to draw conclusions
from observational limits on odd images is to use models of
individual systems that take into account not only the detection
limits but also constraints on the global lens model from the
observed images (e.g., Cohn et al.\ 2001; Mu\~noz et al.\ 2001;
Rusin \& Ma 2001; Norbury et al.\ 2001). Instead, in this
statistical analysis I examine the distribution of odd images
predicted by star+halo models to understand the general trends
(c.f., Wallington \& Narayan 1993).

The upper limits on odd images in the CLASS doubles range from
$\fodd<0.0083$ down to $\fodd<0.0005$; to be very conservative, we
can simply say that no CLASS double has an odd image brighter than
$\fodd=0.01$. In contrast, \reffig{core} shows that star+halo models
predict that more than 20\% of lenses should have odd images with
$\fodd \ge 0.01$; and for the range of parameters allowed by the
number and sizes of lenses, the fraction is more like 30\%. In
other words, star+halo models predict that detectable odd images
should be quite common, in conflict with observations. This result
is not terribly sensitive to the model parameters. It is
surprisingly similar for NFW and Moore models, despite the
differences in the dark matter cusps. The explanation is again
adiabatic contraction: most of the mass in the cores of the galaxies
was pulled in by adiabatic contraction, which affects NFW halos
more strongly than Moore halos and thus tends to reduce the
differences between the two models. The fraction of lenses with
detectable odd images is a strong function of the detection
threshold (see \reffigs{cusp}{corebh}), so the discrepancy
between data and models grows as the upper limit on $\fodd$ is
decreased.

Odd images are sensitive to the density profile at small radii, so
the assumption of a Hernquist model galaxy with a $\rho \propto r^{-1}$
cusp should be examined. Faber et al.\ (1997), Ravindranath et al.\
(2001), and Rest et al.\ (2001) find that the surface brightness
distributions $I(R)$ of early-type galaxies have a range of cusps;
luminous early-type galaxies have cores or shallow cusps
($I \propto R^{-\gamma}$ with $\gamma \lesssim 0.3$, corresponding
to $\rho \propto r^{-\alpha}$ with $\alpha \lesssim 1.3$), while
fainter galaxies have steeper power-law cusps. Hence, the Hernquist
model seems reasonable for the massive galaxies that dominate
lensing.  Still, for completeness I consider models where the
Hernquist galaxy is generalized to an arbitrary cusp using the
density profile $\rho \propto r^{-\alpha} (r_s+r)^{\alpha-4}$.
\reffig{cusp} shows the results for models with $\CC=7.7$ and
$\fcool=0.19$ (a point close to the $\theta$ boundary in
\reffig{clev1}).  Steep cusps suppress odd images, but only if they
are considerably steeper than the luminosity cusps in luminous
early-type galaxies.  As a corollary, steep cusps also make
galaxies more efficient lenses and thus aggravate the discrepancy
between the models and the observed number and sizes of lenses.
In other words, cusps do not provide a very attractive resolution
to the odd image problem.  These conclusions apply to both NFW and
Moore dark matter models.

The simple star+halo models may not be sufficient for this
analysis, because many galaxies are observed to contain central
supermassive black holes (e.g., Magorrian et al.\ 1998; Gebhardt
et al.\ 2000a, 2000b; Ferrarese \& Merritt 2000; Merritt \&
Ferrarese 2001). While black holes have little effect on the
number and sizes of lenses (they barely affect the potential on
kiloparsec scales), they may suppress or even eliminate odd images
(Mao, Witt \& Koopmans 2001). To understand their effects on lens
statistics, I add black holes to the star+halo models, where the
mass of the black hole is determined from the velocity dispersion
of the galaxy using the empirical correlation $M_{bh} = (1.30 \pm
0.36) \times 10^{8}\, M_\odot\, (\sigma/200\,\kms)^{4.72\pm0.36}$
(Ferrarese \& Merritt 2000; Merritt \& Ferrarese 2001).\footnote{The
correlation was derived for nearby galaxies, but for simplicity I
use it at all redshifts.  This approach is conservative if black
holes grow no faster than their surrounding galaxies, as in the
models by Haehnelt \& Kauffmann (2000) to explain the correlation.}
\reffig{corebh} shows that black holes normalized by this relation
have little effect on odd images down to $\fodd \sim 0.001$. Making
the black holes systematically more massive increases the suppression,
but only for the faintest odd images. Black holes must lie off the
Ferrarese \& Merritt relation by at least a factor of 10 in mass
before they begin to affect odd images at the $\fodd=0.01$ level.
These conclusions again apply with almost equal strength to NFW
and Moore models.

\subsection{Discussion}

When compared to the current limits from CLASS, the star+halo
models clearly predict too many detectable odd images. The
discrepancy is not easily resolved by changing the density
profile or invoking supermassive black holes. In other words,
galaxies constructed from CDM mass distributions have central
densities that are too low. This conclusion is consistent with
the result from Rusin \& Ma (2001) that the lack of odd images
requires steep density profiles, $\rho \propto r^{-\alpha}$ with
$\alpha > 1.8$ at 90\% confidence. Given that the measurements of
odd images are only upper limits, the constraints from odd images
can only get stronger --- and perhaps substantially, if more lenses
turn out to have upper limits as strong as $\fodd<0.0005$ in
B~0218+357.

The odd image problem stands in stark contrast to most
observational tests of CDM. Spiral galaxy dynamics are said to
imply that CDM halos are {\it too concentrated\/} to agree with
observed galaxies (e.g., Moore 1994; de Blok et al.\ 2001; Salucci
2001; Weiner et al.\ 2001), although there is still substantial
debate about whether beam smearing affects this conclusion for
low surface brightness galaxies (van den Bosch et al.\ 2000; van
den Bosch \& Swaters 2000). Independent of the dynamical arguments,
the number and sizes of observed lenses lead to a similar conclusion,
as shown in \S 4. Recent interest in modifications to CDM, such as
self-interacting dark matter (e.g., Spergel \& Steinhardt 2000;
see Wandelt et al.\ 2000 for a review), has therefore focused on
making dark matter halos less concentrated. However, the lack of
odd images implies that galaxies built from CDM halos are
{\it not concentrated enough\/}, to a significant degree. If
self-interacting dark matter reduces densities, it would only
exacerbate the odd image problem.

It is clearly important to understand and resolve this paradox.
Perhaps it is a case of comparing different samples. Lensing
intrinsically selects dense, massive galaxies, so most lenses are
elliptical galaxies; by contrast, the dynamical arguments against
CDM halos come primarily from spiral galaxies, and in particular
low surface brightness galaxies. However, the strongest limit on an
odd image actually comes from a lens produced by a face-on spiral
galaxy ($\fodd<0.0005$ for B~0218+357).

More likely, it is a question of scales. Dynamical observations
and observed lensed images probe scales from several kiloparsecs
down to $\sim$0.5 kpc, while odd images probe scales more like
tens of parsecs. The paradox could be resolved if halos have high
densities on $\sim$10 pc scales, low mean densities at $\sim$1 kpc,
and substantial dark matter halos beyond several kiloparsecs. The
star+halo models do not show the required small-scale structure
--- but they are based on a questionable extrapolation of CDM
profiles to scales much smaller than the resolution of numerical
simulations. On such small scales, other effects may be important;
while supermassive black holes appear not to suppress odd images
at the required levels, self-interacting dark matter may be a
mechanism to steepen the central cusp (Burkert 2000; Kochanek \&
White 2000; Moore et al.\ 2000), or simply to concentrate a lot
of dark matter at very small radii (Ostriker 2000). Regardless
of what the correct explanation turns out to be, it is clear that
the odd image problem provides a very interesting probe of galaxy
mass distributions in the inner tens of parsecs.

\section{Conclusions}

Star+halo models of elliptical galaxies are a natural outgrowth of
the Cold Dark Matter paradigm that can reproduce the quasi-isothermal
mass distributions implied by stellar dynamics, X-ray halos, and
gravitational lensing. When the stellar components are fixed by
observed galaxy populations, gravitational lens statistics can be
used to constrain the dark matter components. The observed number
and sizes of lenses place important upper limits on the amount of
dark matter in the inner regions of elliptical galaxies: on average,
dark matter can account for no more than about 33\% of the total
mass inside one effective radius ($R_e$), or about 40\% of the mass
inside $2 R_e$ (95\% confidence upper limits). Lensed images
typically appear at a few effective radii, so the stellar and dark
matter components must be comparably important for
lensing.\footnote{Dark matter mass fractions derived from individual
lenses may differ somewhat from the limits just quoted. The quoted
limits are statistical averages and apply to the mass in spheres,
while lens models give individual masses and involve the mass in
cylinders.}

The dark matter limits are interesting when interpreted in the
context of the CDM paradigm. Galaxies built from CDM mass
distributions have significant amounts of central dark matter, so
they predict that lenses should be more numerous and larger than
observed. A high-density ($\Omega_M=1$) CDM cosmology can therefore
be ruled out at better than 95\% confidence, while a low-density,
flat cosmology ($\Omega_M=0.2$, $\Omega_\Lambda=0.8$) is at best
marginally consistent with the lens data. By implying that CDM
model galaxies have too much mass on kiloparsec scales, lensing
independently supports the evidence against CDM from spiral galaxy
dynamics (e.g., Moore 1994; de Blok et al.\ 2001; Salucci 2001;
Weiner et al.\ 2001). The evidence from lensing is important because
its only uncertainty is the decomposition into stellar and dark
matter components, and lensing extends the argument to elliptical
galaxies. These conclusions may be invalid if simple adiabatic
contraction models do not apply to elliptical galaxies, but the
models appear to agree well with simulated galaxies even in merger
scenarios (Gottbrath 2000).

Lensing offers a unique additional insight into galaxy mass
distributions. Star+halo models predict that a central or ``odd''
image should be detectable in more than 30\% of lenses, but such
images are rarely observed. Equivalently, the upper limits on the
fluxes of odd images for CLASS lenses lead to strong lower limits
on the central densities of lens galaxies, and star+halo models
fail to satisfy the limits. The failure is perhaps not surprising,
because odd images are sensitive to the mass distribution on scales
of tens of parsecs, where simple CDM models may break down. For
example, supermassive black holes, which appear to be common in
the centers of galaxies (e.g., Magorrian et al.\ 1998; Gebhardt
et al.\ 2000a, 2000b; Ferrarese \& Merritt 2000; Merritt \&
Ferrarese 2001), might reconcile the models with the data by
helping to suppress odd images (see Mao et al.\ 2001). However,
star+halo models would need black holes that lie off the observed
black hole--bulge mass correlation by more than a factor of 10 in
order to suppress odd images at the required level. Alternatively,
steeper central cusps could suppress odd images, but only if mass
cusps are considerable steeper than luminosity cusps, only if
steep cusps can survive merger events (which is unlikely if the
progenitors contain black holes; Milosavljevi\'c \& Merritt 2001),
and only at the expense of aggravating the discrepancy in the
number and sizes of lenses.

The evidence from the number and sizes of lenses (and from spiral
galaxy dynamics) implies, then, that CDM mass distributions have
too much mass on kiloparsec scales; and the odd image problem
indicates too little mass on tens of parsec scales. The problem
may be the assumption that the dark matter particles are
collisionless. Allowing dark matter self-interactions can lower
the density on large scales (e.g., Spergel \& Steinhardt 2000;
Burkert 2000; Dav\'e et al.\ 2001), although matching observations
may require substantial fine tuning (e.g., Kochanek \& White 2000;
Moore et al.\ 2000; Yoshida et al.\ 2000). As an intriguing
corollary, Ostriker (2000) proposes that self-interactions could
also increase the dark matter density on very small scales.
Self-interacting dark matter models might therefore provide a way
to reconcile cosmological mass models with real galaxies over a
wide range of spatial scales, although the details remain to be
worked out. In any case, it appears that lensing will be a very
important test of modifications to CDM through its ability to
simultaneously test mass distributions on large scales (via
observed images) and small scales (via limits on, and eventually
detections of, odd images).  These tests based on galaxy-scale
lenses will complement constraints on self-interacting dark matter
from giant arcs produced by cluster lenses (Meneghetti et al.\
2001).

It would be interesting to apply star+halo models to individual
lenses, especially given the statistical limits implying that the
stellar and dark matter components are of comparable importance
in lensing. Comparing the halo/galaxy mass ratios inferred for
individual lenses with the limits from statistics would be an
important test of the models. Using two independent components
would make it possible to to examine whether the stellar and
halo distributions have similar or different ellipticities and
orientations, although the decomposition may not be unique.
Finally, using two components with different ellipticities and/or
orientations would test whether single-component lens models are
sufficient or oversimplified. In particular, star+halo models
might provide an internal reason why most lenses cannot be fit
by a single ellipsoidal mass distribution (e.g., Keeton, Kochanek
\& Seljak 1997), although internal effects may often be smaller
than external tidal perturbations from objects near the lens
galaxy or along the line of sight (e.g., Hogg \& Blandford 1994;
Schechter et al.\ 1997).

The constraints on dark matter from lens statistics can be strengthened
with better data in at least three ways. First, the strongest
constraints come from the distribution of image separations, where
the model predictions are sensitive to the redshift distribution of
the CLASS sample. The CLASS spectroscopic subsample (Marlow et al.\
2000) provides a useful starting point, but the redshift distribution
of the full sample must be better constrained to make the lensing
analysis truly robust. Second, many of the current constraints on
odd images lie at the $\fodd \simeq 0.01$ level (Rusin \& Ma 2001).
If the limits could be improved to the $\fodd \simeq 0.003$ level
or better, they would make the models much more sensitive to
supermassive black holes, and lensing would become a powerful probe
of black holes in distant galaxies out to redshift $z \sim 1$.
Finally, this analysis is based primarily on the sample of lenses
from the CLASS survey, which is the largest existing lens survey
but still has only 18 lenses.  Larger surveys, in particular the
Sloan Digital Sky Survey (SDSS; York et al.\ 2000), should increase
the number of lenses by well over an order of magnitude. The SDSS
lens sample will dramatically improve the constraints from lensing,
provided that selection effects are well understood and that there
is a subsample of radio loud lenses where useful limits on odd
images can be obtained.

\acknowledgements
Acknowledgements: I would like to thank a number of people for
assistance with this project:
Matthias Steinmetz for interesting and helpful discussions;
Vince Eke and Romeel Dav\'e for discussions about CDM and SIDM;
Shude Mao, Martin Norbury, and Ben Wandelt for prompting the
analysis of odd images;
Huan Lin for providing data in advance of publication;
Janice Lee for help sorting through the debate about rotation
curves and beam smearing;
and Chris Kochanek and Ann Zabludoff for comments on the manuscript.
This work has been supported by Steward Observatory.

\appendix

\section{An Analytic Solution of Adiabatic Contraction}

Blumenthal et al.\ (1986) give a simple analytic treatment of
spherical adiabatic contraction that agrees remarkably well with
more detailed numerical simulations. Let $M_i(r_i)$ be the initial
mass profile as a function of the initial radius $r_i$, while
$M_g(r)$ and $M_h(r)$ are the final mass profiles of the galaxy and
halo, respectively. In the Blumenthal et al.\ (1986) prescription,
the three profiles are related by two equations,
\begin{eqnarray}
  & r\,\left[ M_g(r) + M_h(r) \right] = r_i\,M_i(r_i)\,, \label{eq:contract1}\\
  & M_h(r) = (1-\fcool) M_i(r_i)\,, \label{eq:contract2}
\end{eqnarray}
where $\fcool = M_{g,tot}/M_{i,tot}$ is the fraction of the
system's mass contained in baryons that cool to form the galaxy.
(There can be other baryons that remain hot and distributed
throughout the halo, but they do not affect adiabatic contraction.)

This adiabatic contraction formalism has often been applied to the
problem of a disk galaxy in an NFW halo. Rix et al.\ (1997) have
computed adiabatic contraction for elliptical galaxies numerically,
but I find that with a Hernquist model (eq.~\ref{eq:hern} in
\S 2.1) the problem can be solved analytically. Each initial radius
$r_i$ maps to a unique final radius $r$ given by the solution of
the equation
\begin{equation} \label{eq:ri2r}
  \fcool\,r^3 + (r+s_g)^2 \left[ (1-\fcool) r - r_i \right] m_i(r_i) = 0\,,
\end{equation}
which is a cubic polynomial in $r$. Note that I have take the
galaxy scale radius $r_s$ from \refeq{hern} and relabeled it as
$s_g$. Also, $m_i(r_i) = M_i(r_i)/M_{200}$ is the initial mass
profile normalized by the virial mass (the mass inside the virial
radius $r_{200}$). In the limit $r \gg s_g$, \refeq{ri2r} has the
simple asymptotic solution
\begin{equation}
  r = {r_i\,m_i(r_i) \over \fcool + (1-\fcool) m_i(r_i)}\ .
\end{equation}
The full general solution can be also be found analytically,
although it cannot be written quite so compactly. Following
Abramowitz \& Stegun (1981), solve a cubic equation of the form
\begin{equation} \label{eq:cubic}
  z^3 + a_2 z^2 + a_1 z + a_0 = 0
\end{equation}
by defining
\begin{eqnarray}
  p &=& {a_1 a_2 - 3 a_0 \over 6} - {a_2^3 \over 27}\,, \\
  q &=& {a_1 \over 3} - {a_2^2 \over 9}\,, \\
  s_1 &=& \left( p + \sqrt{q^3+p^2} \right)^{1/3} , \\
  s_2 &=& \left( p - \sqrt{q^3+p^2} \right)^{1/3} .
\end{eqnarray}
There is always a real solution of \refeq{cubic} at
\begin{equation}
  z_1 = (s_1 + s_2) - {a_2 \over 3}\ .
\end{equation}
There are two other roots that may be real or complex, but because
the $r_i \to r$ mapping under adiabatic contraction should be
one-to-one, only the single real root is relevant. Once the cubic
equation has been solved to map $r_i$ to $r$, \refeq{contract1} can
be used to write the total mass profile as
\begin{equation} \label{eq:Mtot}
  M_{tot}(r) \equiv M_g(r) + M_h(r) = {r_i \over r}\,M_i(r_i)\,.
\end{equation}
This solution of adiabatic contraction by a Hernquist galaxy can be
used for any form of the initial halo, by simply inserting the
desired initial profile $M_i(r_i)$ into \refeqs{ri2r}{Mtot}.

\section{The Mean Magnification of Odd Images}

The mean magnification of odd images can be computed analytically,
at least for simple lens models. Neglecting magnification bias, the
mean magnification is defined to be
\begin{equation} \label{eq:modd1}
  \avg{\mu_{odd}} = { {\int \mu_{odd}(\uu)\,d\uu} \over
    {\int d\uu} }\ ,
\end{equation}
where the integrals extend over the multiply-imaged region of the
source plane. Changing variables in the numerator to integrate in
the image plane yields
\begin{equation} \label{eq:modd2}
  \avg{\mu_{odd}} = { {\int_{odd} d\xx} \over {\int d\uu} }\ ,
\end{equation}
where the integral in the numerator extends over the region in
the image plane where odd images are found. The result is so simple
because the $\mu_{odd}(\uu)$ factor in \refeq{modd1} is exactly
cancelled by the Jacobian of the coordinate transformation.
\refEq{modd2} says that the mean odd image magnification is simply
the area where odd images occur in the image plane divided by the
area of the multiply-imaged region of the source plane. This result
is general and does not require specific symmetries in the lens
model. It can be generalized to other types of images as well,
provided that multiplicities are properly counted.

Now focusing on spherical systems, $\avg{\mu_{odd}} =
(r_{cr}/u_{out})^2$ where $r_{cr}$ is the radial critical curve
and $u_{out}$ is the boundary of the multiply-imaged region (see
\S 3). These radii are found as follows. The critical radius is
the solution of the equation
\begin{equation}
  {d\phi_{R} \over dR}\biggr|_{R=r_{cr}} = 1 ,
\end{equation}
where $\phi_{R}$ is the lensing deflection. The boundary of the
multiply-imaged region is then
\begin{equation}
  u_{out} = (\phi_{R} - R) \bigr|_{R=r_{cr}} .
\end{equation}
Consider two simple models. First, for a softened isothermal
sphere with density $\rho \propto (s^2+r^2)^{-1}$ with core radius
$s$,
\begin{eqnarray}
  \avg{\mu_{odd}} &=& { {4s} \over (\sqrt{4r_E+s}-3\sqrt{s})^2 }\ ,
    \label{eq:modd3} \\
  &=& {s \over r_E} + 3\left({s \over r_E}\right)^{3/2}
    + {13 \over 2}\,\left({s \over r_E}\right)^2
    + {\cal O}\left({s \over r_E}\right)^{5/2} ,
\end{eqnarray}
where $r_E$ is the Einstein ring radius of the model when the core
radius is zero. Second, for a simple power law density $\rho
\propto r^{-\alpha}$ (with $\alpha>1$ to ensure that the projected
mass distribution is a decreasing function of radius),
\begin{equation}
  \avg{\mu_{odd}} = \cases{
    [(2-\alpha)/(\alpha-1)]^2 & if $1 < \alpha < 2$ \cr
    0 & if $\alpha \ge 2$ \cr
  } \label{eq:modd4}
\end{equation}
Note that for $\alpha \ge 2$ the model does not produce odd images
(the density is singular, so the odd image theorem does not apply;
see \S 5.1), so $\avg{\mu_{odd}} \equiv 0$.



\begin{references}

\reference{}
Abramowitz, M., \& Stegun, I. A. 1981, Handbook of mathematical
functions with formulas, graphs, and mathematical tables
(Washington: National Bureau of Standards)

\reference{}
Barkana, R. 1998, \apj, 502, 531

\reference{}
Bartelmann, M. 1996, \aap, 313, 697

\reference{}
Biggs, A. D., Browne, I. W. A., Helbig, P., Koopmans, L. V. E.,
Wilkinson, P. N., \& Perley, R. A. 1999, \mnras, 304, 349

\reference{}
Binggeli, B., \& Cameron, L. A. 1991, \aap, 252, 27

\reference{}
Binggeli, B., \& Cameron, L. A. 1993, \aaps, 98, 297

\reference{}
Blais-Ouellette, S., Amram, P., \& Carignan, C. 2001, \aj, 121, 1952

\reference{}
Blumenthal, G., Faber, S., Flores, R., \& Primack, J. 1986, \apj,
301, 27

\reference{}
Brighenti, F., \& Mathews, W. G. 1998, \apj, 495, 239

\reference{}
Browne, I. W. A. 2001, in Gravitational Lensing: Recent Progress
and Future Goals (ASP), ed. T. Brainerd \& C. S. Kochanek

\reference{}
Bruzual, G., \& Charlot, S. 1993, \apj, 405, 538

\reference{}
Bullock, J. S., Kolatt, T. S., Sigad, T., Somerville, R. S.,
Kravtsov, A. V., Klypin, A. A., Primack, J. R., \& Dekel, A.
2001, \mnras, 321, 559

\reference{}
Burke, W. L. 1981, \apj, 244, L1

\reference{}
Burkert, A. 2000, \apj, 534, L143

\reference{}
Carroll, S. M., Press, W. H., \& Turner, E. L. 1992, \araa, 30, 499

\reference{}
Chae, K.-H., Khersonsky, V. K., \& Turnshek, D. A. 1998, \apj, 506, 80

\reference{}
Chen, G. H., \& Hewitt, J. N. 1993, \aj, 106, 1719

\reference{}
Cohn, J. D., Kochanek, C. S., McLeod, B. A., \& Keeton, C. R. 2001,
\apj, in press (also preprint astro-ph/0008390)

\reference{}
Cole, S., \& Lacey, C. 1996, \mnras, 281, 716

\reference{}
Crone, M. M., Evrard, A. E., \& Richstone, D. O. 1994, \apj, 434, 402

\reference{}
Dav\'e, R., Spergel, D. N., Steinhardt, P. J., \& Wandelt, B. D.
2001, \apj, 547, 574

\reference{}
Debattista, V. P., \& Sellwood, J. A. 1998, \apj, 493, L5

\reference{}
de Blok, W. J. G., McGaugh, S. S., Bosma, A., \& Rubin, V. C. 2001,
\apj, 552, L23

\reference{}
Djorgovski, S., \& Davis, M. 1987, \apj, 313, 59

\reference{}
Dressler, A., Lynden-Bell, D., Burstein, D., Davies, R. L., Faber, S. M.,
Terlevich, R., \& Wegner, G. 1987, \apj, 313, 42

\reference{}
Drinkwater, M. J., Webster, R. L., Francis, P. J., Condon, J. J.,
Ellison, S. L., Jauncey, D. L., Lovell, J., Peterson, B. A., \&
Savage, A. 1997, \mnras, 284, 85

\reference{}
Dubinski, J. 1994, \apj, 431, 617

\reference{}
Eke, V., Navarro, J. F., \& Steinmetz, M. 2000, preprint (astro-ph/0012337)

\reference{}
Fabbiano, G. 1989, \araa, 27, 87

\reference{}
Falco, E. E., Kochanek, C. S., \& Mu\~noz, J. A. 1998, \apj, 494 47

\reference{}
Ferrarese, L., \& Merritt, D. 2000, \apj, 539, L9

\reference{}
Flores, R. A., Primack, J. R., Blumenthal, G. R., \& Faber, S. M.
1993, \apj, 412, 443

\reference{}
Flores, R. A., \& Primack, J. R. 1994, \apj, 427, L1

\reference{}
Gebhardt, K., et al.\ 2000a, \apj, 539, L13

\reference{}
Gebhardt, K., et al.\ 2000a, \apj, 543, L5

\reference{}
Gottbrath, C. 2000, MS thesis, University of Arizona

\reference{}
Haehnelt, M. G., \& Kauffmann, G. 2000, \mnras, 318, L35

\reference{}
Helbig, P. 2000, preprint (astro-ph/0008197)

\reference{}
Henstock, D. R., Browne, I. W. A., Wilkinson, P. N., \& McMahon, R. G.
1997, \mnras, 290, 380

\reference{}
Hernquist, L. 1990, \apj, 356, 359

\reference{}
Hogg, D. W., \& Blandford, R. D. 1994, \mnras, 268, 889

\reference{}
Huss, A., Jain, B., \& Steinmetz, M. 1999a, \apj, 517, 64

\reference{}
Huss, A., Jain, B., \& Steinmetz, M. 1999b, \mnras, 308, 1011

\reference{}
Ibata, R. A., Lewis, G. F., Irwin, M. J., Leh\'ar, J., \& Totten, E. J.
1999, \aj, 118, 1922

\reference{}
Jing, Y. P., \& Suto, Y. 2000, \apj, 529, L69

\reference{}
Keeton, C. R., Kochanek, C. S., \& Seljak, U. 1997, \apj, 482, 604

\reference{}
Keeton, C. R. 1998, PhD thesis, Harvard University

\reference{}
Keeton, C. R., \& Kochanek, C. S. 1998, \apj, 495, 157

\reference{}
Keeton, C. R., \& Madau, P. 2001, \apj, 549, L25

\reference{}
Keeton, C. R. 2001, preprint (astro-ph/0102341)

\reference{}
Klypin, A., Kravtsov, A. V., Bullock, J. S., \& Primack, J. R. 2000,
preprint (astro-ph/0006343)

\reference{}
Kochanek, C. S. 1993, \apj, 419, 12

\reference{}
Kochanek, C. S. 1995, \apj, 445, 559

\reference{} 
Kochanek, C. S. 1996, \apj, 466, 638

\reference{} 
Kochanek, C. S., Falco, E. E., Impey, C. D., Leh\'ar, J., McLeod, B. A.,
Rix, H.-W., Keeton, C. R., Mu\~noz, J. A., \& Peng, C. Y. 2000, \apj, 543, 131

\reference{}
Kochanek, C. S., Falco, E. E., Impey, C. D., Leh\'ar, J., McLeod, B. A.,
\& Rix, H.-W. 2001, CfA/Arizona Space Telescope Lens Survey web site
({\tt http://cfa-www.harvard.edu/castles})

\reference{}
Kochanek, C. S., \& White, M. 2000, \apj, 543, 514

\reference{}
Kochanek, C. S., \& White, M. 2001, preprint (astro-ph/0102334)

\reference{}
Lin, H., Yee, H. K. C., Carlberg, R. G., Morris, S. L., Sawicki, M.,
Patton, D. R., Wirth, G., \& Shepherd, C. W. 1999, \apj, 518, 533

\reference{}
Lin, H., et al. 2001, in preparation

\reference{}
Loewenstein, M., \& White, R. E. 1999, \apj, 518, 50

\reference{}
Magorrian, J., et al. 1998, \aj, 115, 2285

\reference{}
Marlow, D. R., Rusin, D., Jackson, N., Wilkinson, P. N., Browne, I. W. A.,
\& Koopmans, L. 2000, \aj, 119, 2629

\reference{}
Marlow, D. R., et al. 2001, \aj, 121, 619

\reference{}
Mao, S., Witt, H. J., \& Koopmans, L. V. E. 2001, \mnras, 323, 301

\reference{}
Maoz, D., \& Rix, H.-W. 1993, \apj, 416, 425

\reference{}
McGaugh, S. S., \& de Blok, W. J. G. 1998, \apj, 499, 41

\reference{}
Meneghetti, M., Yoshida, N., Bartelmann, M., Moscardini, L.,
Springel, V., Tormen, G., \& White, S. D. M. 2001, \mnras, 325, 435

\reference{}
Merritt, D., \& Ferrarese, L. 2001, \apj, 547, 140

\reference{}
Milosavljevi\'c, M., \& Merritt, D. 2001, preprint (astro-ph/0103350)

\reference{}
Moore, B. 1994, \nat, 370, 629

\reference{}
Moore, B., Governato, F., Quinn, T., Stadel, J., \& Lake, G. 1998,
\apj, 499, L5

\reference{}
Moore, B., Quinn, T., Governato, F., Stadel, J., \& Lake, G. 1999,
\mnras, 310, 1147

\reference{}
Moore, B., Gelato, S., Jenkins, A., Pearce, F. R., \& Quilis, V.
2000, \apj, 535, L21

\reference{}
Moore, B. 2001, preprint (astro-ph/0103100)

\reference{}
Mu\~noz, J. A., Kochanek, C. S., \& Keeton, C. R. 2001, \apj, in press
(also preprint astro-ph/0103009)

\reference{}
Navarro, J. F., Frenk, C. S., \& White, S. D. M. 1996, \apj, 462, 563

\reference{}
Navarro, J. F., Frenk, C. S., \& White, S. D. M. 1997, \apj, 490, 493

\reference{}
Navarro, J. F., \& Steinmetz, M. 2000, \apj, 528, 607

\reference{}
Norbury, M., et al. 2001, \mnras, submitted

\reference{}
Ostriker, J. P. 2000, \prl, 84, 5258

\reference{}
Phillips, P., et al. 2001, \mnras, submitted

\reference{}
Porciani, C., \& Madau, P. 2000, \apj, 532, 679

\reference{}
Press, W. H., Teukolsky, S. A., Vetterling, W. T., \& Flannery, B. P.
1992, Numerical Recipes in C: The Art of Scientific Computing,
Second Edition (New York: Cambridge Univ. Press)

\reference{}
Ravindranath, S., Hi, L. C., Peng, C. Y., Filippenki, A. V., \&
Sargent, W. L. W. 2001, preprint (astro-ph/0105390)

\reference{}
Rest, A., van den Bosch, F. C., Jaffe, W., Tran, H., Tsvetanov, Z.,
Ford, H. C., Davies, J., \& Schafer, J. 2001, \aj, 121, 2431

\reference{}
Rix, H.-W., de Zeeuw, P. T., Cretton, N., van der Marel, R. P., \&
Carollo, C. M. 1997, \apj, 488, 702

\reference{}
Rusin, D., \& Ma, C.-P. 2001, \apj, 549, L33

\reference{}
Rusin, D., \& Tegmark, M. 2001, \apj, 553, 709

\reference{}
Salucci, P., \& Burkert, A. 2000, \apj, 537, L9

\reference{}
Salucci, P. 2001, \mnras, 320, L1

\reference{}
Schade, D., Barrientos, L. F., \& L\'opez-Cruz, O. 1997, \apj, 477, L17

\reference{}
Schechter, P. L. 1976, \apj, 203, 297

\reference{}
Schechter, P. L., et al. 1997, \apj, 475, L85

\reference{}
Schneider, P., Ehlers, J., \& Falco, E. E. 1992, Gravitational Lenses
(New York: Springer)

\reference{}
Spergel, D. N., \& Steinhardt, P. J. 2000, \prl, 84, 3760

\reference{}
Swaters, R. A., Madore, B. F., \& Trewhella, M. 2000, \apj, 531, L107

\reference{}
Tegmark, M., Zaldarriaga, M., \& Hamilton, A. J. S. 2001, \prd, 63, 43007

\reference{}
Turner, E. L. 1980, \apj, 242, L135

\reference{}
Turner, E. L., Ostriker, J. P., \& Gott, J. R. 1984, \apj, 284, 1

\reference{}
Turner, E. L. 1990, \apj, 365, L43

\reference{}
Tytler, D., O'Meara, J. M., Suzuki, N., \& Lubin, D. 2000, \physscr, 85, 12

\reference{}
van den Bosch, F. C., Robertson, B. E., Dalcanton, J. J., \& de Blok, W. J. G.
2000, \aj, 119, 1579

\reference{}
van den Bosch, F. C., \& Swaters, R. A. 2000, preprint (astro-ph/0006048)

\reference{}
van Dokkum, P. G., Franx, M., Kelson, D. D., \& Illingworth, G. D.
1998, \apj, 504, L17

\reference{}
Wallington, S., \& Narayan, R. 1993, \apj, 403, 517

\reference{}
Wandelt, B. D., Dav\'e, R., Farrar, G. R., McGuire, P. C., Spergel, D. N.,
\& Steinhardt, P. J. 2000, in Proceedings of Dark Matter 2000 (also
preprint astro-ph/0006344)

\reference{}
Weiner, B. J., Sellwood, J. A., \& Williams, T. B. 2001, \apj, 546, 931

\reference{}
White, S. D. M., Navarro, J. F., Evrard, A. E., \& Frenk, C. S. 1993,
\nat, 366, 429

\reference{}
Wyithe, J. S. B., Turner, E. L., \& Spergel, D. N. 2000, preprint
(astro-ph/0007354)

\reference{}
York, D. G., et al. 2000, \aj, 120, 1579

\reference{}
Yoshida, N., Springel, V., White, S. D. M., \& Tormen, G. 2000, \apj, 544, L87

\end{references}
\end{document}